\documentclass[twocolumn]{aastex63}

\received{}
\revised{}
\accepted{}
\submitjournal{ApJ}

\shorttitle{A $\theta_{E}\sim40\arcsec$ Cluster Lens: RXC J0032.1+1808}
\shortauthors{Acebron et al.}

\begin{document}

\title{RELICS: A Very Large ($\theta_{E}\sim40\arcsec$) Cluster Lens -- RXC J0032.1+1808}

\correspondingauthor{Ana Acebron}
\email{anaacebronmunoz@gmail.com}

\author[0000-0003-3108-9039]{Ana Acebron}
\affiliation{Physics Department,
Ben-Gurion University of the Negev, P.O. Box 653,
Be'er-Sheva 8410501, Israel}

\author[0000-0002-0350-4488]{Adi Zitrin}
\affiliation{Physics Department,
Ben-Gurion University of the Negev, P.O. Box 653,
Be'er-Sheva 8410501, Israel}

\author[0000-0001-7410-7669]{Dan Coe}
\affiliation{Space Telescope Science Institute, 3700 San Martin Drive, Baltimore, MD 21218, USA}

\author[0000-0003-3266-2001]{Guillaume Mahler}
\affiliation{Department of Astronomy, University of Michigan, 1085 South University Ave, Ann Arbor, MI 48109, USA}

\author[0000-0002-7559-0864]{Keren Sharon}
\affiliation{Department of Astronomy, University of Michigan, 1085 South University Ave, Ann Arbor, MI 48109, USA}

\author[0000-0002-7559-0864]{Masamune Oguri}
\affiliation{Research Center for the Early Universe, University of Tokyo, Tokyo 113-0033, Japan}
\affiliation{Department of Physics, University of Tokyo, Tokyo 113-0033, Japan}
\affiliation{Kavli Institute for the Physics and Mathematics of the Universe (Kavli IPMU, WPI), University of Tokyo, Chiba 277-8582, Japan}

\author[0000-0001-5984-0395]{Maru\v{s}a Brada\v{c}}
\affiliation{Department of Physics, University of California, Davis, CA 95616, USA}

\author[0000-0002-7908-9284]{Larry D. Bradley}
\affiliation{Space Telescope Science Institute, 3700 San Martin Drive, Baltimore, MD 21218, USA}

\author[0000-0003-1625-8009]{Brenda Frye}
\affiliation{Department of Astronomy, Steward Observatory, University of Arizona, 933 North Cherry Avenue, Tucson, AZ, 85721, USA}

\author{Christine J. Forman}
\affiliation{Harvard-Smithsonian Center for Astrophysics, 60 Garden Street, Cambridge, MA 02138, USA}

\author[0000-0002-6338-7295]{Victoria Strait}
\affiliation{Department of Physics, University of California, Davis, CA 95616, USA}

\author{Yuanyuan Su}
\affiliation{Department of Physics and Astronomy, University of Kentucky, 505 Rose Street, Lexington, KY, 40506, USA}

\author[0000-0002-7196-4822]{Keiichi Umetsu}
\affiliation{Academia Sinica Institute of Astronomy and Astrophysics (ASIAA), No. 1, Section 4, Roosevelt Road, Taipei 10617, Taiwan}

\author[0000-0002-8144-9285]{Felipe Andrade-Santos}
\affiliation{Harvard-Smithsonian Center for Astrophysics, 60 Garden Street, Cambridge, MA 02138, USA}

\author[0000-0001-9364-5577]{Roberto J. Avila}
\affiliation{Space Telescope Science Institute, 3700 San Martin Drive, Baltimore, MD 21218, USA}

\author[0000-0002-3772-0330]{Daniela Carrasco}
\affiliation{School of Physics, University of Melbourne, VIC 3010, Australia}

\author[0000-0002-8261-9098]{Catherine Cerny}
\affiliation{Astronomy Department and Institute for Astrophysical Research, Boston University, 725 Commonwealth Ave., Boston, MA 02215, USA}

\author{Nicole G. Czakon}
\affiliation{Academia Sinica Institute of Astronomy and Astrophysics (ASIAA), No. 1, Section 4, Roosevelt Road, Taipei 10617, Taiwan}

\author[0000-0003-0248-6123]{William A. Dawson}
\affiliation{Lawrence Livermore National Laboratory, P.O. Box 808 L-210, Livermore, CA, 94551, USA}

\author{Carter Fox}
\affiliation{Department of Astronomy, University of Michigan, 1085 South University Ave, Ann Arbor, MI 48109, USA}

\author[0000-0001-8989-2567]{Austin T. Hoag}
\affiliation{Department of Physics, University of California, Davis, CA 95616, USA}

\author[0000-0001-7826-6448]{Kuang-Han Huang}
\affiliation{Department of Physics, University of California, Davis, CA 95616, USA}

\author[0000-0002-8829-5303]{Traci L. Johnson}
\affiliation{Department of Astronomy, University of Michigan, 1085 South University Ave, Ann Arbor, MI 48109, USA}

\author[0000-0003-2449-6314]{Shotaro Kikuchihara}
\affiliation{Institute for Cosmic Ray Research, The University of Tokyo, 5-1-5 Kashiwanoha, Kashiwa, Chiba 277-8582, Japan}
\affiliation{Department of Astronomy, Graduate School of Science, The University of Tokyo, 7-3-1 Hongo, Bunkyo, Tokyo, 113-0033, Japan}

\author[0000-0002-6536-5575]{Daniel Lam}
\affiliation{Leiden Observatory, Leiden University, NL-2300 RA Leiden, The Netherlands}

\author[0000-0002-3754-2415]{Lorenzo Lovisari}
\affiliation{Harvard-Smithsonian Center for Astrophysics, 60 Garden Street, Cambridge, MA 02138, USA}

\author[0000-0003-0094-6827]{Ramesh Mainali}
\affiliation{Department of Astronomy, Steward Observatory, University of Arizona, 933 North Cherry Avenue, Tucson, AZ, 85721, USA}

\author[0000-0001-6342-9662]{Mario Nonino}
\affiliation{INAF - Osservatorio Astronomico di Trieste, Via Tiepolo 11, I-34131 Trieste, Italy}

\author[0000-0001-5851-6649]{Pascal A. Oesch}
\affiliation{Department of Astronomy, University of Geneva, Chemin des Maillettes 51, 1290 Versoix, Switzerland}

\author{Sara Ogaz}
\affiliation{Space Telescope Science Institute, 3700 San Martin Drive, Baltimore, MD 21218, USA}

\author[0000-0002-1049-6658]{Masami Ouchi}
\affiliation{Institute for Cosmic Ray Research, The University of Tokyo, 5-1-5 Kashiwanoha, Kashiwa, Chiba 277-8582, Japan}
\affiliation{Kavli Institute for the Physics and Mathematics of the Universe (Kavli IPMU, WPI), University of Tokyo, Kashiwa, Chiba 277-8583, Japan}

\author{Matthew Past}
\affiliation{Department of Astronomy, University of Michigan, 1085 South University Ave, Ann Arbor, MI 48109, USA}

\author[0000-0003-3653-3741]{Rachel Paterno-Mahler}
\affiliation{Department of Astronomy, University of Michigan, 1085 South University Ave, Ann Arbor, MI 48109, USA}

\author{Avery Peterson}
\affiliation{Department of Astronomy, University of Michigan, 1085 South University Ave, Ann Arbor, MI 48109, USA}

\author[0000-0003-0894-1588]{Russell E. Ryan}
\affiliation{Space Telescope Science Institute, 3700 San Martin Drive, Baltimore, MD 21218, USA}

\author[0000-0002-7453-7279]{Brett Salmon}
\affiliation{Space Telescope Science Institute, 3700 San Martin Drive, Baltimore, MD 21218, USA}

\author{Daniel P. Stark}
\affiliation{Department of Astronomy, Steward Observatory, University of Arizona, 933 North Cherry Avenue, Tucson, AZ, 85721, USA}

\author[0000-0003-3631-7176]{Sune Toft}
\affiliation{Cosmic Dawn Center (DAWN)}
\affiliation{Niels Bohr Institute, University of Copenhagen, Lyngbyvej 2, DK-2100, Copenhagen, Denmark}

\author[0000-0001-9391-305X]{Michele Trenti}
\affiliation{School of Physics, University of Melbourne, VIC 3010, Australia}
\affiliation{Australian Research Council, Centre of Excellence for All Sky Astrophysics in 3 Dimensions (ASTRO 3D), Melbourne VIC, Australia}

\author[0000-0003-0980-1499]{Benedetta Vulcani}
\affiliation{INAF-Osservatorio Astronomico di Padova, Vicolo Dell’osservatorio 5, 35122 Padova Italy}

\author{Brian Welch}
\affiliation{Department of Physics and Astronomy, The Johns Hopkins University, 3400 North Charles Street, Baltimore, MD 21218, USA}



\begin{abstract}
Extensive surveys with the \textit{Hubble Space Telescope} (HST) over the past decade, targeting some of the most massive clusters in the sky, have uncovered dozens of galaxy-cluster strong lenses. The massive cluster strong-lens scale is typically $\theta_{E}\sim10\arcsec$ to $\sim30-35\arcsec$, with only a handful of clusters known with Einstein radii $\theta_{E}\sim40\arcsec$ or above (for  $z_{source}=2$, nominally). Here we report another very large cluster lens, RXC J0032.1+1808 ($z=0.3956$), the second richest cluster in the redMapper cluster catalog and the 85th most massive cluster in the Planck Sunyaev-Zel'dovich catalog.
With our Light-Traces-Mass and fully parametric (dPIEeNFW) approaches, we construct strong lensing models based on 18 multiple images of 5 background galaxies newly identified in the \textit{Hubble} data mainly from the \textit{Reionization Lensing Cluster Survey} (RELICS), in addition to a known sextuply imaged system in this cluster.
Furthermore, we compare these models to Lenstool and GLAFIC models that were produced independently as part of the RELICS program.
All models reveal a large effective Einstein radius of $\theta_{E}\simeq40\arcsec$ ($z_{source}=2$), owing to the obvious concentration of substructures near the cluster center. Although RXC J0032.1+1808 has a very large critical area and high lensing strength, only three magnified high-redshift candidates are found within the field targeted by RELICS. Nevertheless, we expect many more high-redshift candidates will be seen in wider and deeper observations with \textit{Hubble} or \emph{JWST}. Finally, the comparison between several algorithms demonstrates that the total error budget is largely dominated by systematic uncertainties.

\end{abstract}



\section{Introduction} \label{sec:intro}
Over the past decade, extensive galaxy cluster lensing campaigns have been undertaken with the \emph{Hubble Space Telescope} \citep[HST;][]{Postman2012, Schmidt2014, Treu2015, Lotz2017, Coe2019,Steinhardt2020arXiv}. \textit{HST}'s unique combination of sensitivity and resolution allows for the identification of lensed galaxies that are multiply imaged by the targeted galaxy clusters \citep[see for instance][]{Franx1997, Frye1998, Broadhurst2005, Diego2018, Jauzac2019, Caminha2019, Lagattuta2019}. These multiple images, in turn, allow us to construct mass models for the clusters, describing the underlying matter distribution. 
Most (albeit not all) of the clusters targeted with \emph{HST}, which are typically estimated to be massive based on X-ray, the Sunyaev-Zel’dovich effect \citep[SZ,][]{Sunyaev1970}, or optical richness criteria (and lensing signatures such as giant arcs in ground-based data), show multiply imaged background
galaxies in numbers that generally increase with strong lens scale \citep[or critical area, i.e. the area enclosed by the critical curves of infinite magnification; see for instance][]{Vega-Ferrero2019}. 

\begin{deluxetable*}{c c c c c c c}
\tablecaption{\label{tab:RE}
    Known strong lensing clusters with Einstein radius above $\theta_{E} > 35"$ ($z_{s}=2$)$\dagger$.}
\tablehead{Galaxy cluster name & R.A. & Decl& $\mathrm{z}$& $\theta_{E}$\tablenotemark{a} & 
Surveys\tablenotemark{b} & References\tablenotemark{c} \\
&[J2000]&[J2000]&&& &}
\startdata
MACS J0717.5+3745 & 07:17:34& +37:44:49& $0.5460$ & $\sim55\arcsec$ &HFF\tablenotemark{d}, CLASH\tablenotemark{e}, BUFFALO\tablenotemark{f} & \citet{Zitrin2009b}\\
Abell 1689 & 13:11:34& -01:21:56& $0.1890$ &$\sim45\arcsec$ &-& \citet{Broadhurst2005}\\
PLCK G287.0+32.9 & 11:50:49& -28:05:07& $0.3800$& $\sim42\arcsec$ &RELICS\tablenotemark{g}& \citet{Zitrin2017}\\
RXCJ2211.7-0349 & 22:11:43& -03:49:45& $0.3970$& $\sim 41\arcsec$ & RELICS& \citet{Cerny2018}\\
Abell 370 &02:39:53 &-01:34:36 & $0.3750$ & $\sim39\arcsec$ &HFF, BUFFALO& \citet{Richard2010}\\
RXC J0032.1+1808 &00:32:11 &+18:07:49 & $0.3956$ & $\sim 40\arcsec$& RELICS& This work \\
PLCK G171.9-40.7 &03:12:57 & +08:22:19 & $0.2700$ & $\sim 37\arcsec$ &RELICS& \citet{Acebron2018}\\
RCS2 J232727.6-020437 & 23:27:08 & -02:04:54 & $0.6986$& $\sim 35\arcsec$&RELICS&  \citet{Sharon2015}\\
\hline
\enddata
\tablenotetext{a}{The Einstein radii are obtained from strong lensing analyses using different algorithms.}
\tablenotetext{b}{Recent \textit{HST} lensing surveys that included the cluster. }
\tablenotetext{c}{First strong-lensing analysis to publish the size of the lens. More references are available in the literature for some of the clusters.}
\tablenotetext{d}{\textit{Hubble Frontier Fields} Survey; see \citet{Lotz2017}}
\tablenotetext{e}{\textit{ Cluster Lensing And Supernova Survey with Hubble}; see \citet{Postman2012}}
\tablenotetext{f}{\textit{Beyond Ultra-deep Frontier Fields and Legacy Observations} Survey; see \citet{Steinhardt2020arXiv}}
\tablenotetext{g}{\citet{Coe2019}}
\tablenotetext{\dagger}{We note that there are other clusters that were claimed to have large Einstein radii, but later downsized in updated analyses. For example, both MACS J2129.4-0741 and MACS J0257.1-2325 were reported by \citet{Zitrin2011} to have Einstein radii above 35\arcsec, yet an updated analysis with additional broadband data (\citealt{Zitrin2015}, and \emph{in preparation}, respectively) suggests these are smaller lenses. As another example, RX J1347-1145 was analyzed by \citet{Halkola2008} to have an Einstein radius above 35\arcsec as well. An updated analysis using CLASH data by \citet{Zitrin2015}, resulted in a somewhat smaller value of $\theta_{E}\sim33\arcsec$ for a redshift $z_s=2.0$. }
\end{deluxetable*}

Following the high projected mass densities in their centres, the strong-lens scale of galaxy clusters is expected to reach $\theta_{E}$ of the order of tens of arcseconds, with $\theta_{E}$ being the effective Einstein radius (i.e., the radius of the area enclosed by the critical curves of the lens, were it a circle). This is indeed the range of typical Einstein radii found in lensing analyses of well-known clusters \citep{Richard2010b, Oguri2012, Zitrin2015, Sharon2020}. 
Since the Einstein radius size essentially depends on the mass enclosed in the core of the cluster, the distribution of Einstein radii can be used to probe cosmological models as well as structure formation and evolution in its framework \citep{Turner1984, Narayan1988, Oguri2009}. Due to the shape of the cosmic mass function \citep{Tinker2008}, more massive clusters and thus, generally, larger Einstein radii become rarer \citep[see for instance][]{Oguri2012, Zitrin2012}. Other effects are also known to boost the lensing cross-section such as a major-axis orientated along the line-of-sight, a high degree of substructure, cluster concentration or the dynamical state  \citep{Meneghetti2010}.
Indeed, only a handful of clusters are known to have Einstein radii of $\theta_{E} \gtrsim 35\arcsec$, for a source at $z_{s}\sim2$, nominally (see the list in Table \ref{tab:RE}). 
The number of clusters with particularly large Einstein radii is therefore very important because it can place useful constraints on structure formation and evolution models \citep[e.g.][]{Oguri2009}. 

Larger strong lenses (presenting larger Einstein radii) should have on average larger areas of high magnification and thus lens more background sources \citep{Vega-Ferrero2019}, especially if the faint-end slope of the luminosity function is steep enough \citep{Broadhurst1995, Bradley2014, Coe2015}, as is currently estimated, for example, for high-redshift galaxies \citep{Bouwens2015, Finkelstein2015, Mason2015}. Building a census of this particular class of large cluster lenses is thus important also for efficiently searching for high-redshift galaxies with current and future observations.

The \textit{Reionization Lensing Cluster Survey} \citep[RELICS; PI: Coe;][]{Coe2019} is a large \emph{Hubble Space Telescope} program that has observed 41 galaxy clusters chosen largely based on SZ-mass estimates from the Planck PSZ2 all-sky catalog \citep{Planck2016}. One of the chief goals of the RELICS survey is to identify bright high-redshift galaxies that could be followed up from the ground and with the upcoming \textit{James Webb Space Telescope} (\textit{JWST}). Lens models for the observed clusters are needed to study the dark matter distribution, as well as the intrinsic properties of newly uncovered high-redshift galaxy candidates lensed by these clusters \citep{Salmon2020, Salmon2018}. 

In our systematic analysis of RELICS clusters \citep{Cerny2018, Acebron2018, Acebron2019, Cibirka2018, Paterno-Mahler2018, Mahler2019}, we have analyzed RXC J0032.1+1808\footnote{Also known as PSZ1 G116.48-44.47, WHL J8.03426+18.10 and MACS J0032.1+1808.} \citep{Bohringer2001,Ebeling2001}, located at $\mathrm{R.A.=00h32m11.0s}$, $\mathrm{Decl=+18d07m49.0s}$ at a redshift of $z=0.3956$. RXC J0032.1+1808 (RXC0032 hereafter) is the second richest galaxy cluster in the SDSS DR8 redMaPPer cluster catalog \citep[only after RMJ224319.8-093530.9; see][]{Rykoff2014}, but only the 85th most massive cluster in the PSZ2 catalog \citep{Planck2016}, with a mass of $\mathrm{M_{500}}=7.61^{+0.57}_{-0.63}\times 10^{14} \mathrm{M_{\odot}}$, where $M_{500}$ is defined as the cluster mass within the radius $R_{500}$ inside which the mean mass-density is 500 times the critical density $\delta_c$. 

A first strong-lensing (SL) model for this cluster, prior to RELICS imaging, was used in \citet{Dessauges2017} to estimate the properties of a notable multiply imaged system at a redshift of $z=3.6314$ (system 1 \& 2 here). Here we present our strong lensing (SL) analysis of RXC0032 in RELICS data, revealing a very large critical area, similar to only a handful of other clusters known to date (see Table \ref{tab:RE}). We report here this discovery and detail our SL modeling.

The outline of the paper is as follows: in Section \ref{sec:OBS} we briefly describe the data and observations used to identify multiple images for the SL analysis, which is presented in Section \ref{sec:SLF}. The results are shown and discussed in Section \ref{sec:RES}. We compare our results with those obtained from the Lenstool and GLAFIC pipelines\footnote{The Lenstool and GLAFIC models are also publicly available through MAST\textsuperscript{\ref{mast}}.} in Section \ref{sec:COMP} before summarizing our work in Section \ref{sec:DIS}. Throughout we assume a $\Lambda$CDM cosmology with $\Omega_{\rm m0}=0.3$, $\Omega_{\Lambda 0}=0.7$, $H_{0}=70$ km s$^{-1}$Mpc$^{-1}$ where $1\arcsec= 5.337$ kpc at the redshift of RXC J0032.1+1808.

\section{Data and Observations} \label{sec:OBS}

\begin{figure*}
    \centering
    \includegraphics[width=\linewidth]{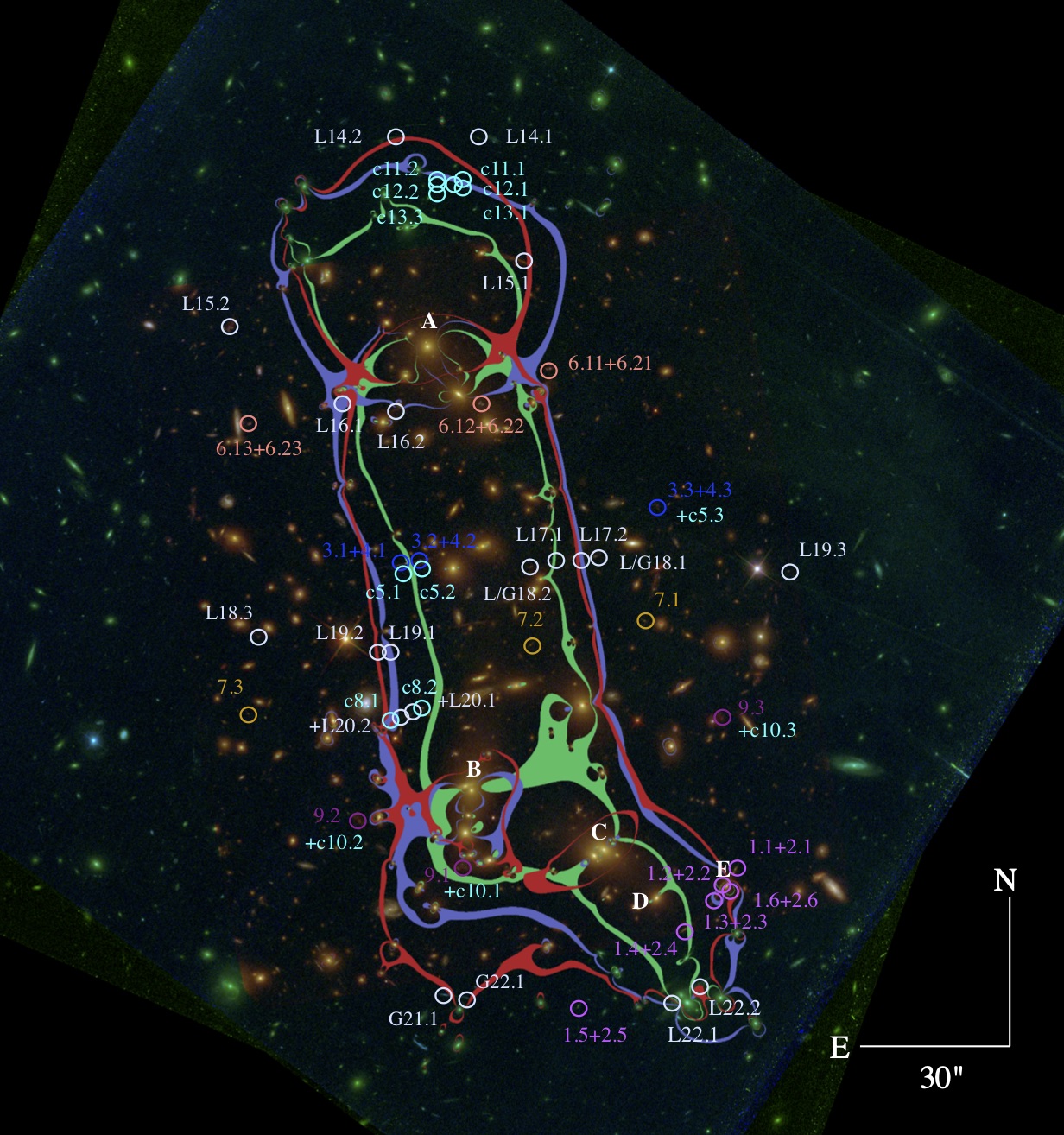} 
    \caption{Color-composite RGB image of RXC0032. The image was constructed with the \textit{HST/ACS} passbands F435W (blue), a combination of F606W+F814W (green), and a combination of the \textit{HST/WFC3IR} passbands F105W+F125W+F140W+F160W (red). The resulting critical curves from our best-fit LTM model are displayed for a source at $z=2.0$ (in green) and $z=9.0$ (in violet). In addition, we add the critical lines at  $z=9.0$ from our fully parametric best-fit dPIEeNFW model (in red). Multiple images (color-coded to ease their identification) are numbered according to Table ~\ref{tab:MI}. Cyan systems are reported as candidates and not used in the LTM/dPIEeNFW models. Systems reported by the Lenstool and/or GLAFIC modeling teams are indicated in light grey.
    Labeled galaxies indicate the cluster members whose weight is freely optimized in the LTM model. The positions of the large-scale dark matter halos for the dPIEeNFW model are fixed to the position of the galaxies A, B and C. The ellipticities and position angles of the two cluster galaxies, B and C, are freely optimized in both models. Finally, we choose the galaxy `A' as the center for the computation of the convergence profile in Figure \ref{kappa}.}
    \label{cc}
\end{figure*}

RXC0032 is part of the RELICS cluster sample \citep{Coe2019}. Each cluster field in the RELICS program was observed for 2 orbits with \textit{Wide Field Camera 3} (WFC3/IR) in the F105W, F125W, F140W, and F160W bands,
and complemented archival observations with the \textit{Advanced Camera for Surveys} (ACS)
so that each field is observed for 3 orbits -- one in each of the passbands F435W, F606W, and F814W.
In addition, 30 hours per band in each of the \textit{Spitzer}-IRAC channels (PI: M. Bradac, PI: Soifer) were observed.
As one orbit of \textit{HST} archival observations already existed for RXC0032 (program GO 12166, PI: Ebeling), RELICS observed this galaxy cluster for only 2 orbits with each of the ACS bands, in addition to the two WFC3 orbits \citep{Coe2019}.

Data reduction of the HST images is described in \citet{Coe2019}. We used the photometric source catalogs generated with Source Extractor \citep{Bertin1996} in dual-image mode from the final drizzled 0.06"/pixel images. The photometric redshifts we use (hereafter $\mathrm{z_{phot}}$), were derived using the \textit{Bayesian Photometric Redshift} program \citep[BPZ,][]{Benitez2000, Benitez2004, Coe2006} from seven \textit{HST} band imaging-data (from the combined RELICS and the aforementioned archival \textit{HST} data). 
The reduced imaging, catalogs, and data products are available for the community through the Mikulski Archive for Space Telescopes (MAST)\footnote{\url{https://archive.stsci.edu/prepds/relics/}\label{mast}}.

\section{Strong lensing formalism} \label{sec:SLF}

We construct lens models for RXC0032 using two methodologies. Primarily, we use the Light-Traces-mass (LTM) methodology, outlined in \citet{Zitrin2015} (and references therein); see also \citet{Broadhurst2005}. With this methodology, we uncover multiple image sets and publish the first estimate for the size of the lens. We also use for comparison a fully parametric formalism, dubbed hereafter dPIEeNFW. Both methodologies assume two main mass components for the mass distribution: one that accounts for the cluster galaxies, and a second that represents the dark matter distribution. While member galaxies are also represented differently, the main difference between the two methodologies is that in the LTM methodology the dark matter distribution is assumed to follow the light distribution, whereas in the parametric model it is independent and follows a combination of symmetric, analytic forms.
Both methodologies are implemented on a grid where the resolution can be changed for computational time purposes. Both methodologies are implemented in the same pipeline \citep{Zitrin2015}, and are briefly detailed below.

\subsection{LTM} \label{ltm}
The LTM formalism is based on the assumption that the cluster member luminosity-weighted distribution is a reasonable tracer for the dark matter component in the cluster. The first ingredient for the model are the cluster galaxies, in which each galaxy is assigned a power-law surface density mass distribution, scaled by its luminosity. The exponent is the same for all galaxies and the superposition of all galaxy contributions constitutes the member-galaxies component of the model. This map is then smoothed with a Gaussian kernel to represent the DM distribution component. The two components are then added with a relative weight and scaled to match a multiple image system (or redshift) of choice. This basic model includes only four free parameters: the power-law exponent for the mass distribution of the cluster galaxies, the width of the smoothing Gaussian, the galaxy to dark matter (DM) weight, and the overall normalization. In addition, we typically include a two-parameter external shear to add further flexibility (manifested mainly in the form of ellipticity of the critical curves), bringing the number of free parameters to six. It is also possible to leave some galaxy masses to be independently scaled  - especially for brighter cluster galaxies, for which we typically find the mass to light (M/L) ratio to be a few times higher than that of other cluster members. These key galaxies can also be assigned an ellipticity, and a core. Finally, the redshifts of systems with no spectroscopic redshift can be optimized by the lens model.

\begin{figure}
    \centering
    \includegraphics[width=\columnwidth]{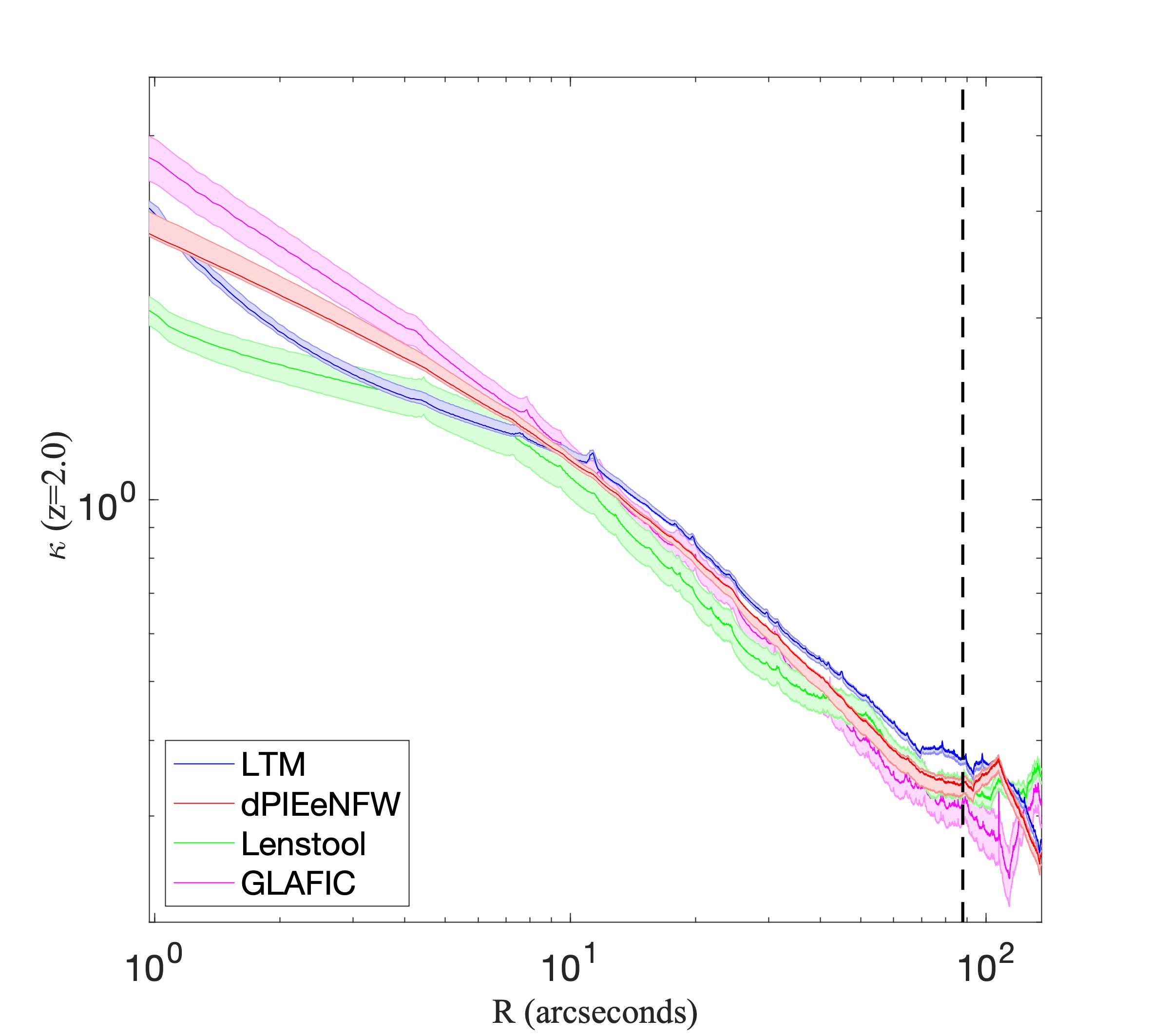}
    \caption{Convergence profiles from our best-fit LTM and dPIEeNFW models, scaled to a source redshift of $z=2.0$. Also plotted for comparison are the Lenstool and GLAFIC models (see Section \ref{sec:COMP}). All profiles are obtained by radially averaging the convergence map in radial annuli, centered on R.A=8.049482; Decl=18.143662, indicated with a `A' in Figure \ref{cc}. The vertical dashed line indicates the radius within which multiple images have been identified.}
    \label{kappa}
\end{figure} 

\subsection{dPIEeNFW}
Our parametric formalism is based on analytic functions for both the member-galaxies and the DM components.
Galaxy-scale halos are parametrized each as a double pseudo-isothermal elliptical (dPIE) mass distribution \footnote{In previous analyses with this pipeline the naming used for the galaxy component was PIEMD, which stands for a \emph{pseudo-isothermal elliptical mass distribution}; the dPIE is a combination of two PIEMDs -- which is what we incorporate in practice.} \citep[see for instance][]{Eliasdottir2007}. Their velocity dispersion, core radius and cut-off radius are scaled based on their luminosity following common scaling relations. These are found to describe well early-type galaxies and can be defined with respect to a typical reference luminosity of a galaxy at the cluster's redshift \citep{Jullo2007, Monna2015, Bergamini2019}. All cluster galaxies, aside from the few brightest cluster galaxies in the cluster core, have no ellipticities assigned to them and their positional parameters are fixed to those derived from their light distribution. 
Each large-scale DM halo is represented by an elliptical Navarro Frenk and White mass distribution \citep[eNFW,][]{Navarro1996} where their concentration, mass, ellipticity, and position angle are free parameters of the model. The central positions of the NFW halos can also be freely optimized but here they are fixed to the light centroid of the brightest cluster galaxies. 

\subsection{Minimization}

For both methodologies, the best-fitting model parameters are found by minimizing the distance in the image-plane between the observed and model-predicted positions of the multiple-image centres, via a $\chi^2$ criterion \citep[the equations for the $\chi^2$ and RMS calculations are presented in][]{Acebron2019}.
To do so, we use a Monte Carlo Markov Chain (MCMC) engine with a Metropolis-Hastings algorithm that typically includes several thousand steps after the burn-in phase, from which both the best-fit model and the uncertainties are derived. For more details on the modeling scheme see \citet{Zitrin2015} and references therein.

\begin{figure*}
    \centering
    \includegraphics[width=\columnwidth]{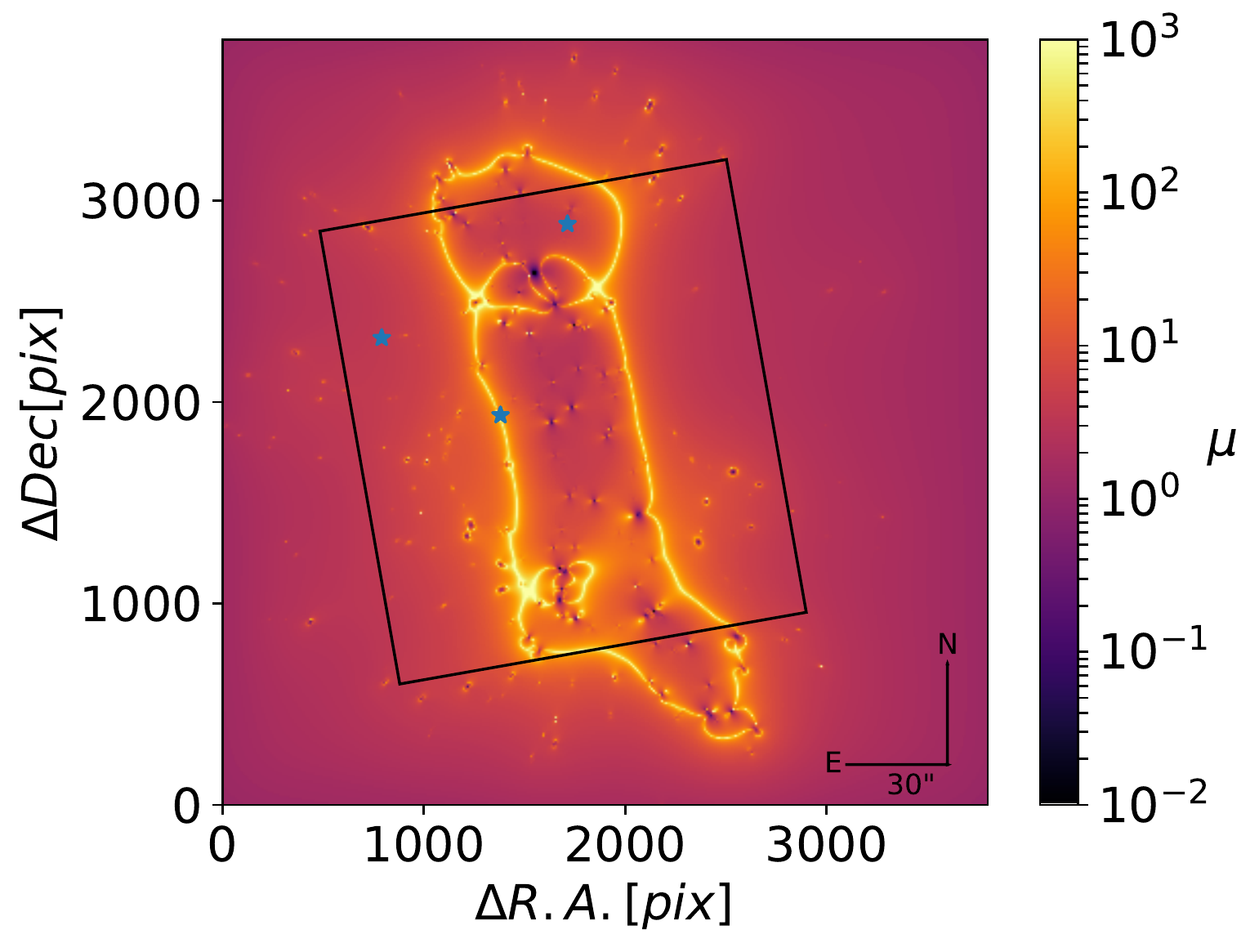}
    \includegraphics[width=1.1\columnwidth]{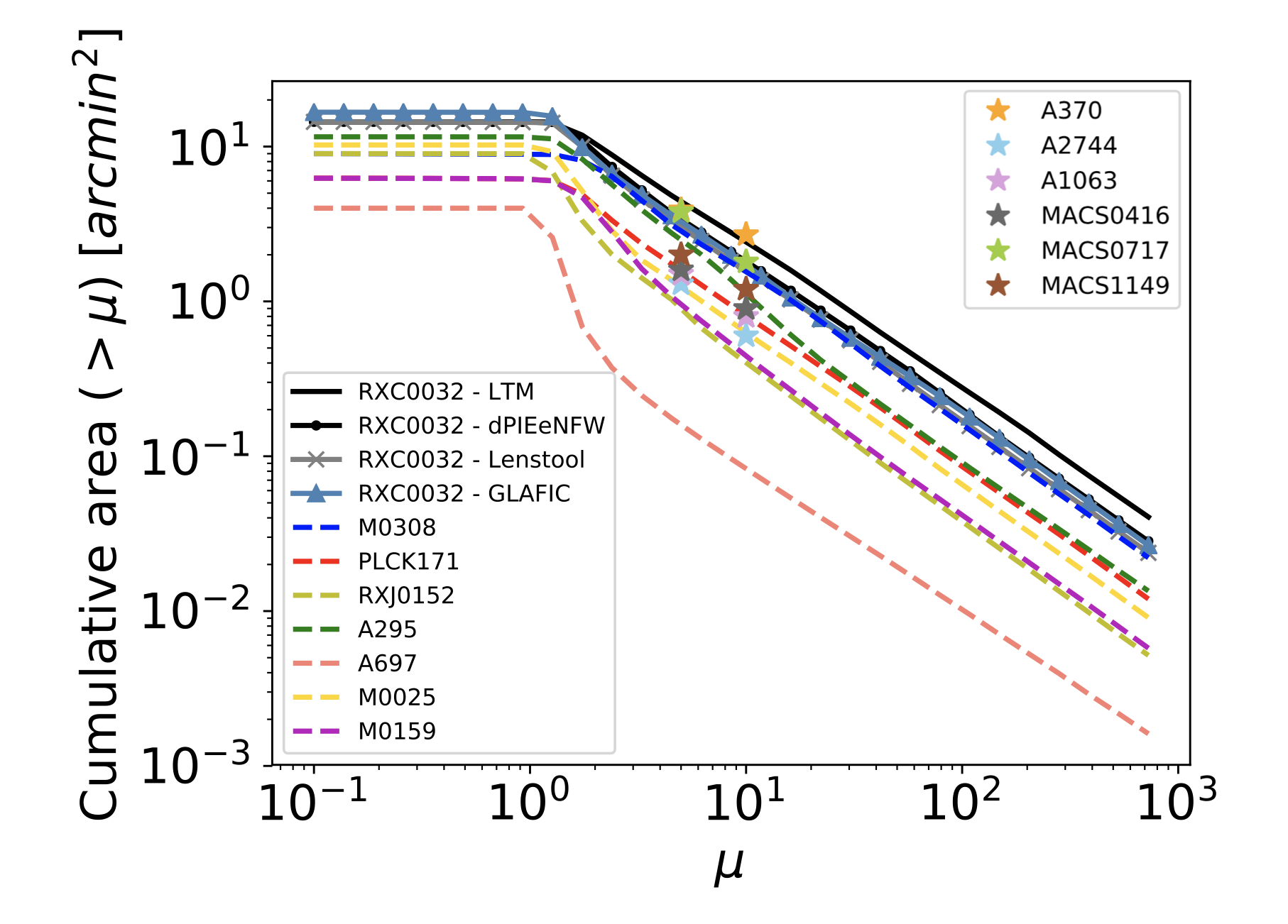}
    \caption{\textbf{Left panel:} Magnification map from our best-fit LTM model for a source at $z_s=6$ in which the RELICS high-z candidates found by \citet{Salmon2020} are marked as blue stars. The black rectangle indicates the WFC3/IR FOV. \\
    \textbf{Right panel:} Cumulative area having a magnification higher than a given value for a source at $z_s=9.0$ in RXC0032's field from our LTM and dPIEeNFW, Lenstool and GLAFIC best fit models. For comparison, we also show the lensing strengths from other RELICS SL clusters modeled with the LTM pipeline, MACS J0308.9+2645, PLCK G171.9-40.7 and Abell S295, Abell 697, MACS J0025.4-1222, MACS J0159.8-0849 and RXJ0152.7-135. The cumulative areas ($\mu>5$ and $\mu>10$) for the \textit{Hubble Frontier Fields} clusters are also indicated as colored stars, computed from the submitted \textsc{zitrin-ltm-gauss} models. The $1\sigma$ errors are typically of the size of the star symbol. It should be noted that different clusters have been modeled with different fields-of-view. For example, the $\mu>5$ and $\mu>10$ values presented for the HFF cluster MACSJ0717.5+3745 are likely be underestimated since its critical curves extend beyond the field-of-view of the HFF observations.} 
    \label{mag}
\end{figure*} 

\subsection{Strong lensing analysis of RXC J0032.1+1808} \label{SLanalysis}

The starting point of our modeling relies on the construction of a cluster member catalog based on the red-sequence method \citep{Gladders2000}. 
We use the magnitudes measured from the F606W and F814W filters to draw a color-magnitude diagram. We only consider galaxies down to 24 AB within $\pm0.3$ mag of the sequence \citep{DeLucia2008}. To exclude stars from our selection we do not include objects whose magnitudes are brighter than 17 AB or have a \textit{stellarity} index below $<0.95$. 
In addition, we take advantage of the delivered photometric catalog by RELICS to check that all selected cluster members were within $z_{phot} \pm 0.1$ of the mean redshift of the cluster (as measured by the BPZ software).
We finally perform a visual inspection of the selected cluster members. This allows us to discard further interloping galaxies (bright foreground galaxies for instance) or artifacts (such as faint and diffuse objects or double detections), or add missing galaxies that appear to be cluster members based on their colors (not selected initially due to the strict magnitude cuts applied).

For the minimization, we consider (for both models) a positional uncertainty of $1.4\arcsec$. This value has been found to encompass both the underlying statistical uncertainties, the systematic uncertainties between our LTM and dPIEeNFW methods, as well as possible uncertainties arising from structure along the line of sight \citep{Host2012}. To speed up the runtime of the models, and while RELICS \textit{HST} images have a resolution of $0.06\arcsec$/pix, we adopt a resolution of $0.24\arcsec$/pixel for the minimization.\\

We present now the multiply imaged systems used in this work, labeled in Table \ref{tab:MI} and Figure \ref{cc}.
We first include the multiply imaged system reported by \citet{Dessauges2017} who identified six two-knot images belonging to this lensed object. We label this as systems 1 and 2 in Table \ref{tab:MI} and Figure \ref{cc}, corresponding to the two different emission knots).\\
In addition, based on the full \textit{HST} \textit{ACS} + \textit{WFC3/IR} dataset, now including RELICS imaging, we detect additional multiply imaged systems (without a spectroscopic confirmation yet), which we identify based on their morphology and color similarity, as well as predictions from a preliminary LTM model. These images are included as lensing constraints in our final models and we briefly review them here.
Systems 3, 4 and c5 straddle the critical curve each forming two bright knots stretching into an arc with their counter-images on the opposite side of the cluster (we note however that the identification of the counter image remains less secure; the LTM, dPIEeNFW and GLAFIC models implement this image in the strong lensing modeling while the Lenstool model does not). 
System 6 is lensed into three images appearing as two distinct pink and blue knots (systems 6.1 and 6.2 in our modeling) in the composite \textit{HST} \textit{ACS} and \textit{WCF3/IR} images (Figure~\ref{cc}), making this identification reliable.
System 7 is lensed into three images which are identified mainly thanks to their similar, red dropout colors as they do not present any peculiar morphology. This system is a relatively high-redshift dropout object ($z_{phot}\sim4.41$ based on its first image for which the photometric redshift is the most reliable), hence its redshift is assumed correct and we fix it in our model. 
Finally, system 9 is lensed into three diffuse red images (in the \textit{ACS} and \textit{WFC3} composite image). 
We also identify additional, potential systems that were however not included in our SL modeling as constraints. 
System c8 appears to be lensed into two images straddling the critical curve. We do not, however, identify a third counter-image on the other side of the critical curves so we report it only as a candidate system.
System c10 lies next to system 9 and is comprised of two faint emission knots. Our SL models predict additional counter-images near a group of cluster member galaxies located at R.A=8.0470889 deg; Decl=18.117521 deg. However, due to the light contamination, we cannot make a reliable identification. Systems c11, c12 and c13 consist each of two arcs straddling the critical curve in the northern region of the cluster. Lacking a \textit{WFC3/IR} coverage in that northern area of the cluster, and BPZ yielding significantly different photometric redshift estimates we decided to keep these systems as candidate systems. Spectroscopic confirmation of these systems would help to more accurately constrain the mass distribution in the most northern region, where no other lensing constraints are seen.

\begin{itemize}
    \item \textbf{LTM}:

The LTM model is built by considering the weight of the 5 brightest cluster members (labeled in Figure \ref{cc}) as free parameters, i.e. allowing their mass-to-light (M/L) ratio to vary. We also consider as free parameters the ellipticity (varying within a flat, small prior of $\pm0.05$ from the measured value from the light distribution) and position angle (varying within $\pm5^{\circ}$ from the input value we measured) of the bright galaxies located at R.A=8.04691, Decl=18.118922; and R.A=8.039185, Decl=18.115616, (galaxies B and C in Figure \ref{cc}).\\
We scale our model to the spectroscopic redshift of systems 1 and 2 (see Table~\ref{tab:MI}). The redshift of the remaining systems, except dropout system 7, are left as free parameters to be optimized in the minimization procedure (allowing the relative $\mathrm{D_{LS}}/\mathrm{D_S}$ ratio for each system, corresponding to its best-fit $z_{phot}$ value, to vary by up to $\pm0.05$; with $\mathrm{D_{S}}$ and $\mathrm{D_{LS}}$ being the angular diameter distances to the source and between the lens and the source).\\
Taking into account the additional freely optimized cluster members and source redshifts, our final model includes a total of 20 free parameters. The resulting critical curves (for a source at $z_s = 2$ and $z_s = 9$) for our final best-fit model which has an image reproduction $RMS =1.60\arcsec$, are shown in Figure~\ref{cc}.

    \item \textbf{dPIEeNFW}:
In merging clusters such as the one analyzed here, multiple DM halos are usually incorporated in the modeling in order to better explain the mass distribution. Our best-fit model comprises three large-scale DM halos whose centres are fixed to the position of the cluster members labeled A, B and C in Figure \ref{cc}. The  large-scale  DM  halos are parametrized with elliptical NFWs, where their ellipticity parameters, concentration and mass are optimized. Cluster members are modeled with a dPIE profile with a fixed core radius of 0.2 kpc for the reference galaxy, a velocity dispersion that is allowed to vary between 80 and 120 km/s, a cut-off radius varying from 45 kpc to 65 kpc. They are modeled as spherical with a mag0 = -21.56 \citep[a reference magnitude for the scaling relations;][]{Faber1976, Jullo2007}.
Similarly as in the LTM model, we leave the ellipticities and position angles of the two bright galaxies presented above to be optimized.\\
We adopt the same multiple images and cluster member catalogs as for the LTM model, leaving as free parameters the redshifts of background sources that have not been spectroscopically confirmed (except for system 7). Our final model includes a total of 24 free parameters.
Our best-fit model has an image reproduction of $RMS =1.43\arcsec$ and for comparison, we also show in Figure \ref{cc} the resulting critical curves for a source at $z_s =9$.

\end{itemize}

\begin{deluxetable*}{c c c c c c c c c c c}
\tablecaption{\label{tab:HZ}
    High-z ($z \sim 6$) lensed candidates}
\tablehead{Galaxy ID\tablenotemark{a} & R.A. & Decl& $\mathrm{J_{125}}$ \tablenotemark{b}& $\mathrm{z_{phot}^{BPZ}}$ \tablenotemark{c} & $\mathrm{z_{phot}^{EZ}}$ \tablenotemark{c}& $\mu^{\mathrm{LTM}}$ \tablenotemark{d}& $\mu^{\mathrm{dPIEeNFW}}$ \tablenotemark{d}& $\mu^{\mathrm{Lenstool}}$ \tablenotemark{d}& $\mu^{\mathrm{GLAFIC}}$ \tablenotemark{d}& $M_{uv,1500}$\tablenotemark{e}\\  
        &[deg]&[deg]&&  &&&&}
\startdata
RXC0032+18-0571  & 8.052543 & 18.131884  &$ 26.42\pm0.25 $ &$6.4^{+0.7}_{-0.8}$&$6.7^{+0.7}_{-1.5}$& $88.43^{+21.89}_{-22.99}$& $19.96^{+4.55}_{-4.40}$&$4.27^{+1.25}_{-1.18}$&$53.20^{+12.49}_{-61.73}$&$-16.10^{+0.45}_{-0.60}$\\
RXC0032+18-0052  &8.046714& 18.147703  &$26.07 \pm 0.26$ & $5.5^{+0.2}_{-0.7}$ & $5.8^{+0.2}_{-0.8}$& $4.82^{+0.17}_{-0.23}$ & $5.17^{+0.35}_{-0.24}$ &$1.82^{+0.18}_{-0.17}$&$2.74^{+0.38}_{-0.37}$&$-18.96^{+0.30}_{-0.35}$\\ 
RXC0032+18-0355 &8.062891 &18.138279   &$26.64 \pm 0.28$ &$5.8^{+0.3}_{-5.1}$&$6.1^{+0.3}_{-5.3}$& $3.75^{+0.08}_{-0.08}$& $2.81^{+0.09}_{-0.12}$ &$2.08^{+0.71}_{-0.81}$&$2.30^{+0.25}_{-0.22}$&$-18.60^{+0.30}_{-0.90}$\\
\hline
\enddata
\tablenotetext{a}{Galaxy ID, following \citet{Salmon2020} notations.}
\tablenotetext{b}{Apparent magnitude in the F125W band.}
\tablenotetext{c}{Redshift estimation based on the BPZ and EAZY pipelines along with their $1\sigma$ uncertainties.}
\tablenotetext{d}{Best-fit magnification estimates (at the respective source redshift) from the LTM, dPIEeNFW, Lenstool and GLAFIC models. The statistical uncertainty is computed as the standard deviation from 100 MCMC models. The LTM best-fit value is the one used for all relevant computations.}
\tablenotetext{e}{Absolute magnitude, $M_{uv}$, at $\lambda = 1500$ \AA\ for which the errors have been propagated from the photometric and magnification uncertainties based on our best-fit LTM model. The resulting rest-frame UV luminosities (corrected for lensing magnifications) have a mean of Muv $\sim$ -17.90 with a standard deviation of 1.56.}
\end{deluxetable*}

\section{Results and discussion} \label{sec:RES}
Figure \ref{cc} shows the critical curves from our lens models. Both the LTM and dPIEeNFW models, despite having a very different representation for the different mass components, yield overall similar critical curves. This is perhaps somewhat expected, given that similar sets of multiple images were used as constraints, although notable differences exist as well - especially in regions of high magnification or regions with fewer constraints. 

Figure \ref{kappa} shows the convergence profile for RXC0032. The SL region is dominated by a large number of substructures, accounting for the shallow inner profile. We also note that the resulting mass distribution has a high overall elongation, or ellipticity, computed as $e=(a^2 - b^2)/ (a^2 + b^2)$, of $e\sim0.74(0.81)\pm0.03$ from the LTM(dPIEeNFW) models. 

We compute the value of the effective Einstein radius as $\theta_E=\sqrt[]{A/\pi}$, with $A$ defined as the area enclosed within the critical curves.
Our strong lensing analysis reveals a particularly prominent lens with a resulting Einstein radii of $\theta_E(z_s=2)\sim38.50\arcsec(39.60\arcsec)\pm 0.20 \arcsec$ and $\theta_E(z_s=9)\sim48.10\arcsec(48.10\arcsec)\pm 0.25\arcsec$ from our LTM(dPIEeNFW) best-fit models, corresponding to an enclosed mass of $M(<\theta_E)=2.13(2.67)\pm 0.2 \times10^{14} M_{\odot}$ within the $z_s=2$ critical curves.
The high-degree of substructures aggregated in the center yields this very large Einstein radius, similar to only a few other clusters.
While the SZ signal, and gas probes in general, lead to the discovery of clusters of high virial mass \citep{Williamson2011, Planck2016}, a high total mass does not guarantee a large strong lensing region, and in the case of RXC0032, it seems that the high cluster richness, for example, better traces the large Einstein radius. RXC0032 portrays the loose relation -- or at least the large scatter in the relation -- between gas probes (and the SZ effect in particular) and the central strong lensing area, which depends more closely on various other factors \citep{Giocoli2016} and especially the amount of matter concentrated or projected in the very center, the cluster triaxial shape and its orientation along the line of sight \citep{King2007, Oguri2009}.

We show in Figure \ref{mag} the $z_s=9$ magnification map from our LTM best-fit model as well as the location of the detected high-redshift galaxy candidates within RXC0032's field-of-view covered by both ACS and WFC3 \citep{Salmon2020}. We compare in Table \ref{tab:HZ} the magnification estimates from both methods and provide the intrinsic $M_{uv}$ at $\lambda=1500\AA$ for the high-redshift candidates.

The derived lensing strength (in the right panel of Figure \ref{mag}) shows that RXC0032 is a very prominent lens, with a large area of high magnification of about $\sim4.4(3.4)$ $\mathrm{arcmin}^2$ with $\mu > 5$ to $\sim2.4(1.8)$ $\mathrm{arcmin}^2$ with $\mu > 10$ from our LTM(dPIEeNFW) best-fit models. RXC0032's lensing strength is significantly larger than that of other RELICS clusters modeled with the LTM pipeline \citep{Acebron2018, Acebron2019, Cibirka2018} and similar or higher than those provided by most \textit{Hubble Frontier Fields} (HFF) cluster lenses.

However, the field of RXC0032 seems to be a unique sight-line, compared to similarly-strong cluster lenses. Despite its lensing strength and large critical area, RXC0032 reveals only three $z\sim6$ high-redshift galaxy candidates detected in the field \citep{Salmon2020}; these are characterized in Table \ref{tab:HZ}. RXJ0152.7-135 \citep[RXJ0152 hereafter;][]{Acebron2019} constitutes an interesting counter-example. The strong lens modeling of both clusters, following the distribution of their member-galaxies, reveals two clusters with similar morphologies, i.e., very elongated and showing a high degree of substructure. These effects have been shown to significantly boost the cluster total cross-section \citep{Meneghetti2007}. This is clearly evident in the case of RXJ0152, which, despite being a much smaller lens ($\theta_{E}(z_{s}=2)\sim9\arcsec$; equivalent to a critical area of 0.06 arcmin$^2$), lenses 24 high-redshift galaxy candidates.\\
The field of RXC0032 provides the lowest yield of high-redshift candidates in comparison to those listed in the right panel of Figure \ref{mag}. While such a small high-redshift candidate sample can be attributed to cosmic variance \citep{Somerville2004, Trenti2008} especially as RELICS targets the brightest distant objects, the low number of high-redshift candidates could also be explained in part by the fact that the \textit{HST/WFC3IR}'s field-of-view ($123\arcsec \times 137 \arcsec$) is fairly small compared to the size of the lens. In Figure \ref{mag} (left panel) we show that a significant proportion of high-magnification regions fall outside of the instrument's field-of-view. Another effect that may play a role in the low number of uncovered high-redshift galaxies is the lower completeness found around the critical curves \citep{Oesch2015}, and conceivably, within them where the intracluster light is brighter. Future deeper observations in a wider field-of-view around RXC0032 should be able to detect more high redshift galaxies as well as to examine these hypotheses.\\
The high-z galaxy candidate RXC0032+18-0571 (see Table \ref{tab:HZ}) lies in a very high-magnification area. In the composite (\textit{ACS}+\textit{WFC3}) image, this object appears as two distinct light emitting knots, one of which appears to be stretched into an $\sim1.5\arcsec$ arc. This is in agreement with our lens model, which predicts a similar stretching for the arc, further supporting the high-z nature of this object. Our SL models also predict a counter-image on the other side of the cluster. However, based on the RELICS high-z photometric study \citep{Salmon2020}, all apparent counter-image candidates have a lower, best-fit photometric redshift estimate. We show the predicted position of the counterimage in Figure \ref{cccomp} based on our LTM model.
Our lens model also indicates that the candidates RXC0032+18-0052 and RXC0032+18-0355 could be two images from the same background source, supporting the high-z nature of these objects as well.

\begin{deluxetable}{c c c}
\tablecaption{\label{tab:RE_rxc0032}
    Effective Einstein radius for RXC J0032.1+1808}
\tablehead{Pipeline & $\theta_E(z_s=2.0)$\tablenotemark{a} & $\theta_E(z_s=9.0)$\tablenotemark{a}\\
& [\arcsec]&[\arcsec]}
\startdata
LTM  & $38.50\pm0.24$  & $48.10\pm0.25$ \\ 
dPIEeNFW  & $39.60\pm0.20$ & $48.10\pm0.22$ \\ 
Lenstool & $37.45\pm0.17$ & $46.41\pm0.23$\\
GLAFIC & $42.20\pm0.70$ & $50.00\pm0.80$\\
\hline
\enddata
\tablenotetext{a}{We note that the errors only represent the statistical uncertainty (computed from 100 random models). However, as seen also here, the systematic uncertainty is typically found to be $\sim10\%$ which better reflects the scatter found between different methods.}
\end{deluxetable}

\section{Comparison}\label{sec:COMP}
As part of the RELICS survey, RXCJ0032 has also been modeled with the Lenstool \citep{Jullo2007} and GLAFIC \citep{Oguri2010, Kawamata2016} pipelines whose high-end products are publicly available through the MAST archive\textsuperscript{\ref{mast}}. In this Section, we compare these two models to our modeling results. We give here a short summary of these two models and refer the reader to the references mentioned above for further details. 

It should be noted that often in comparison studies, the same constraints are used throughout with the goal of comparing the different methodologies explicitly \citep[e.g.,][]{Zitrin2015, Meneghetti2017}. In contrast, here our goal is mainly to probe the credibility of our results and especially, the large Einstein radius estimation. We therefore incorporate the Lenstool and GLAFIC models as well, since these were constructed completely independently by other groups within the RELICS collaboration, including independently identified multiple image sets (presented in Table \ref{tab:MI}). The differences between the results of these different methods also provide the reader with a more quantitative assessment of the magnitude of underlying systematic uncertainties in the presented analysis. It should also be mentioned that while the image recovery RMS is often used in assessing the reliability of strong lens models, a comparison of the RMS values of models that use different constraints is of little use \citep[e.g.,][]{Johnson2016}. 

We first review additional multiple images that were identified independently by the Lenstool (L) and GLAFIC (G) teams and were incorporated in their modeling. System L14 is composed of two multiple images straddling the critical curves in the very northern part of the cluster. However, due to a large RMS value for image L14.2, we would consider this a candidate identification.
The two multiple images of system L15 appear as orange in the composite image and L15.1 is stretched into an arc and may consist of two merging counter images. System L16 is lensed into two diffuse arcs, the second being a radial arc. Systems L17 and L18, each composed of three multiple images, appear respectively as brown and pink emission knots in the central region of the cluster. 
System L19 comprises three multiple images, each presenting a similar morphology with two emission knots.  System L20 lies next to system c8, and is composed of two diffused knots. System G21 is stretched into a blue arc in the composite image with two bright emission knots at each end, and is the only system identified in the southern-east region of this cluster. Finally, system L22 is the most southern multiply imaged system identified and lies next to a small group of cluster members which may introduce a more local galaxy-galaxy lensing effect. To avoid confusion, all multiple images presented throughout this work are shown in Figure \ref{cc} and summarized in Table \ref{tab:MI}. The Table also indicates which images were used as constraints in each model.

We review hereafter the Lenstool and GLAFIC models.

\begin{figure}
    \centering
    \includegraphics[width=\columnwidth]{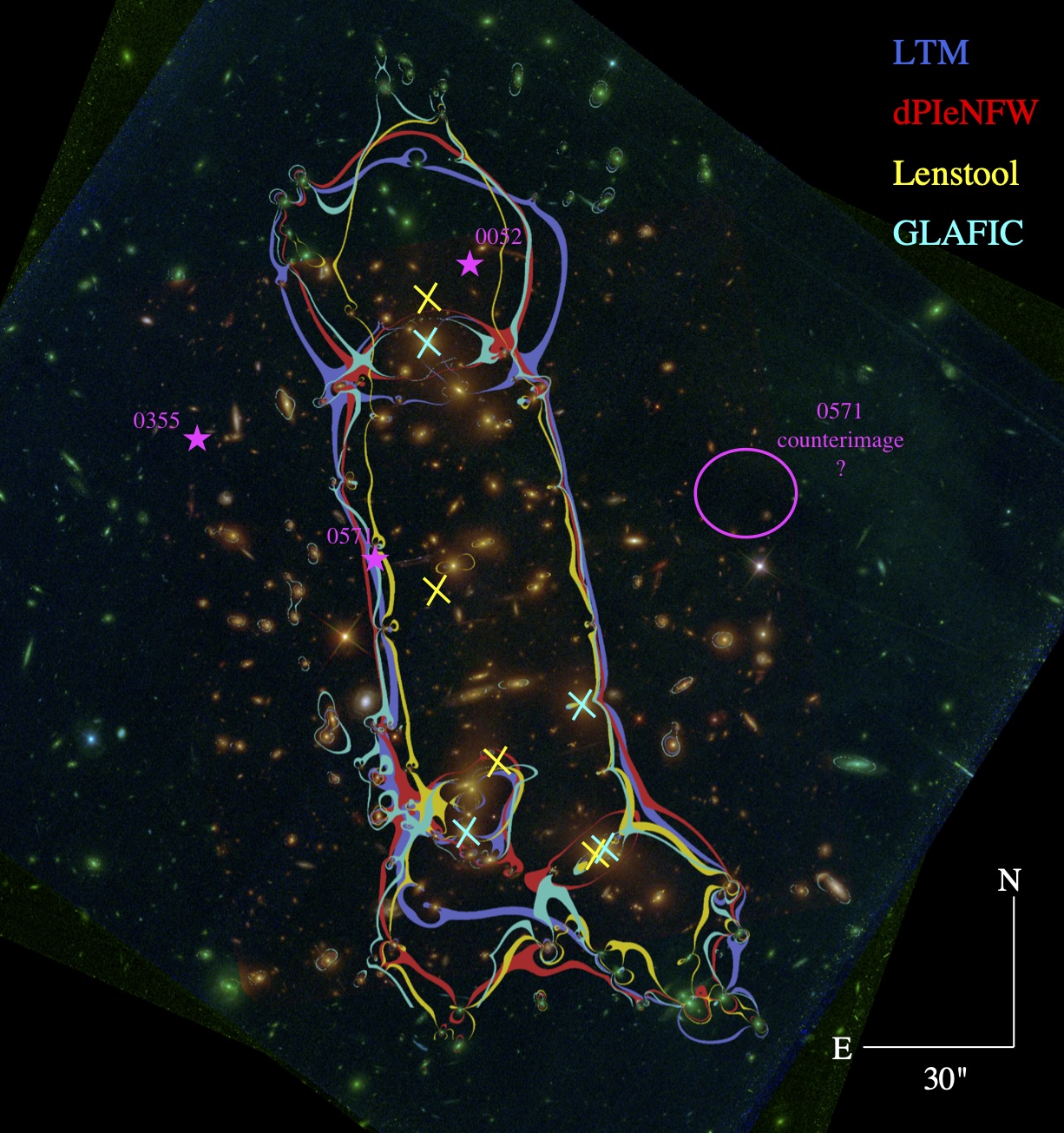}
    \caption{Comparison of the resulting critical curves from the best-fit LTM, dPIEeNFW, Lenstool and GLAFIC models for a source at $z=9.0$ on a color-composite image of RXC0032. The yellow and cyan crosses show the best-fit positions of the large-scale DM halos for the Lenstool and GLAFIC models, respectively. The largest discrepancies occur in the regions where no or different constraints are used in the presented models. The magenta stars indicate the location of the high-z galaxy candidates, and the magenta circle shows the possible location of the counter-image for RXC0032+18-0571 based on the redshifts derived by \citet{Salmon2020}.}
    \label{cccomp}
\end{figure}

\begin{itemize}
    \item \textbf{Lenstool}:
This model is built with 4 large-scale halos parametrised with a dPIE density profile. Their central coordinates, as well as their ellipticity, position angle, core radius and velocity dispersion, are left to be freely optimized. The best fit positions for the large-scale dark matter haloes are found to be: R. A, Decl= (8.048919, 18.1298051); (8.045599, 18.120679); (8.049396, 18.146159); (8.040309, 18.115001), and are shown in Figure \ref{cccomp}.
The small scale haloes associated to galaxy members, identified via the red-sequence method, are modeled with a dPIE profile with a fixed core radius of 0.15 kpc, while both the velocity dispersion and the cut-off radius of a fiducial reference galaxy are freely optimised following the adopted scaling relations \citep{Faber1976, Jullo2007}. Aside from systems 1 and 2 that have a spectroscopic redshift measurement, the redshifts of all other multiple images used in the modeling are freely optimized.
All multiple images are included in the models with a positional uncertainty of $0.3\arcsec$ and the optimization is performed in the image plane. The best-fit model results in an image reproduction of $RMS=0.59\arcsec$.

    \item \textbf{GLAFIC}:
This model includes 4 elliptical NFW large-scale halos which have been fixed to the following coordinates (displayed in Figure \ref{cccomp}): R.A, Decl= (8.049478, 18.143654); (8.047251, 18.116557); (8.039177, 18.115615); (8.040402, 18.123657) while the mass, ellipticity, position angle, and concentration parameters are left as free parameters.
The SL model also includes cluster members identified with the red-sequence method that are modeled as pseudo-Jaffe ellipsoids \citep{Keeton2001} and following the scaling relations. 
The redshifts of all multiple images used in the GLAFIC SL model, but systems 1 and 2, are optimized assuming a Gaussian prior (with $\delta_z=0.5$) around their photo-z estimates. A positional uncertainty of $0.6\arcsec$ is assumed for all multiple images.
The GLAFIC best-fit model has an image reproduction of $RMS=0.49\arcsec$.
\end{itemize}

As can be seen (see also Figures \ref{cc} and \ref{cccomp}), the parametric models (i.e. dPIEeNFW, Lenstool and GLAFIC) have each followed a different modeling prescription: the dPIEeNFW has incorporated three large-scale haloes; the Lenstool and GLAFIC both have four haloes, although their positions slightly vary. The LTM, in contrast, is only semi-parametric and adopts a substantially different methodology. In addition, unlike the other parametric models, the dPIEeNFW is constructed on a fixed grid.

We compare here the main outputs of all four SL models, i.e., the convergence profile, magnification estimates, the resulting critical curves, Einstein radius and the redshift estimates for the multiple images. 

We show in Figure \ref{kappa} a comparison of the convergence profiles. All convergence profiles are in good agreement within the $1\sigma$ error bars. However, the innermost region of the cluster is less well constrained, possibly due to uncertainties related to the chosen modeling techniques. Another issue is that the statistical uncertainties in the LTM and dPIEeNFW models seem to be smaller than those in the Lenstool and GLAFIC models. We are in the process of examining the origin for this discrepancy, which might be, for instance, related to the finite and lower resolution of the LTM and dPIEeNFW models (which also tends to boost the official $\chi^2$ quoted for them). In addition, our fixing of system 7 the highest redshift system used here, to its best photometric redshift value, can also contribute to lowering the errors compared to the two other models. 

The resulting critical curves are found to be in good agreement between all models as shown in Figure \ref{cccomp}. More significant differences are found in regions with no SL constraints such as the most northern or south-eastern regions of the cluster. In the case of the LTM lens model, the differences in the modeling of the southern east region of the cluster may explain the higher RMS value reported for the images 1.5 and 2.5. 

We find that all modeling tools consistently reveal a large Einstein radius (see Table \ref{tab:RE_rxc0032}), with a mean and standard deviation estimates of $\theta_E(z_s=2.0)=39.44\pm2.0$ and $\theta_E(z_s=9.0)=48.15\pm1.47$. 

All models do also estimate RXC0032 to have a prominent lensing strength and are in very good agreement regarding the total area with high magnification (see Figure \ref{mag} - right panel). However, as expected, large discrepancies between reconstructions appear around the lens critical lines, or the highest magnification regions \citep{Meneghetti2017}, as is demonstrated by the magnification estimates of the high-z candidates in Table \ref{tab:HZ}. These values should thus be used with caution.

\begin{figure*}
    \centering
    \includegraphics[width=0.32\linewidth]{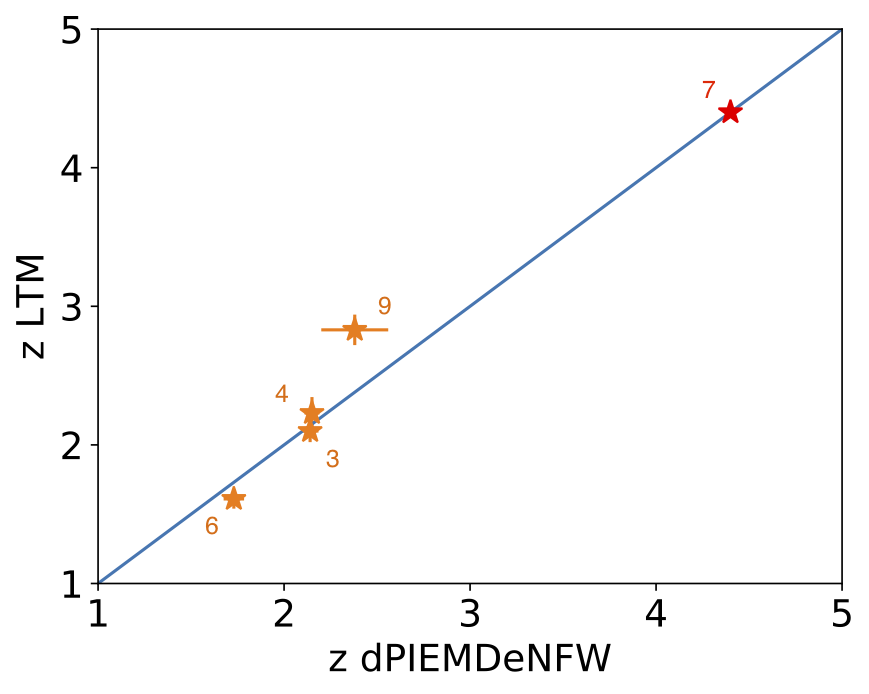}
    \includegraphics[width=0.32\linewidth]{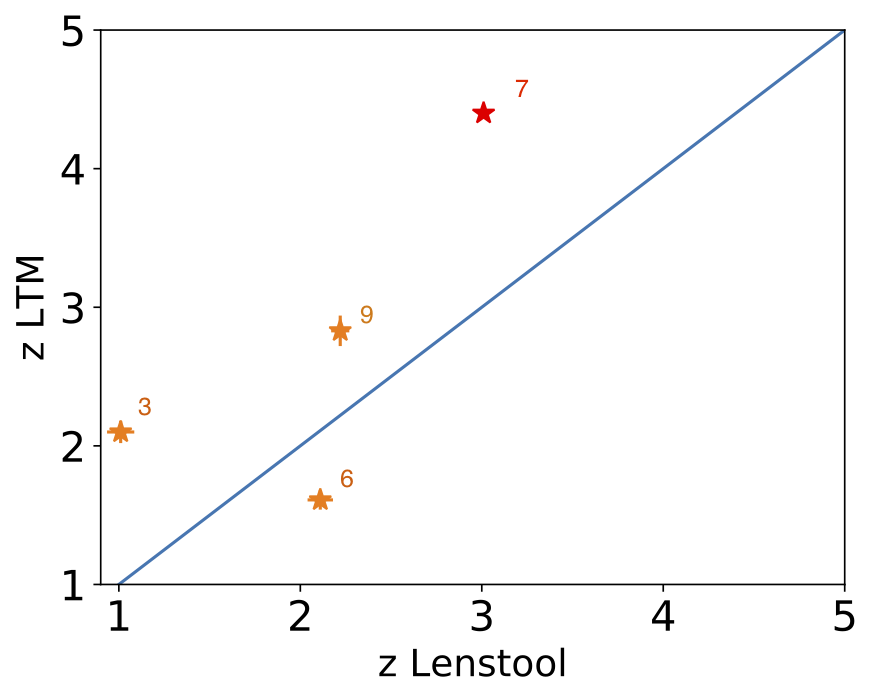}
    \includegraphics[width=0.32\linewidth]{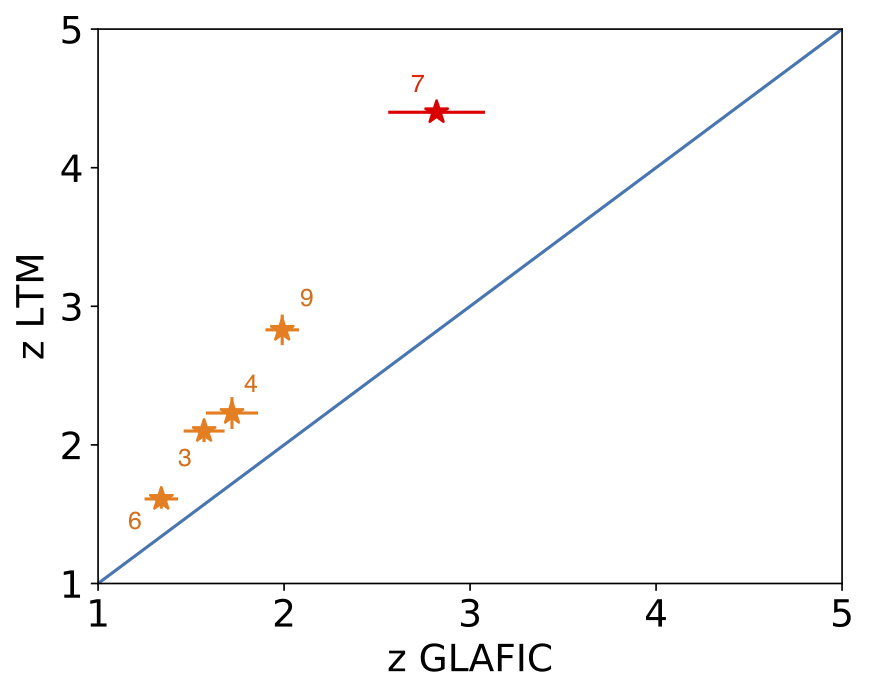}\\
     \includegraphics[width=0.32\linewidth]{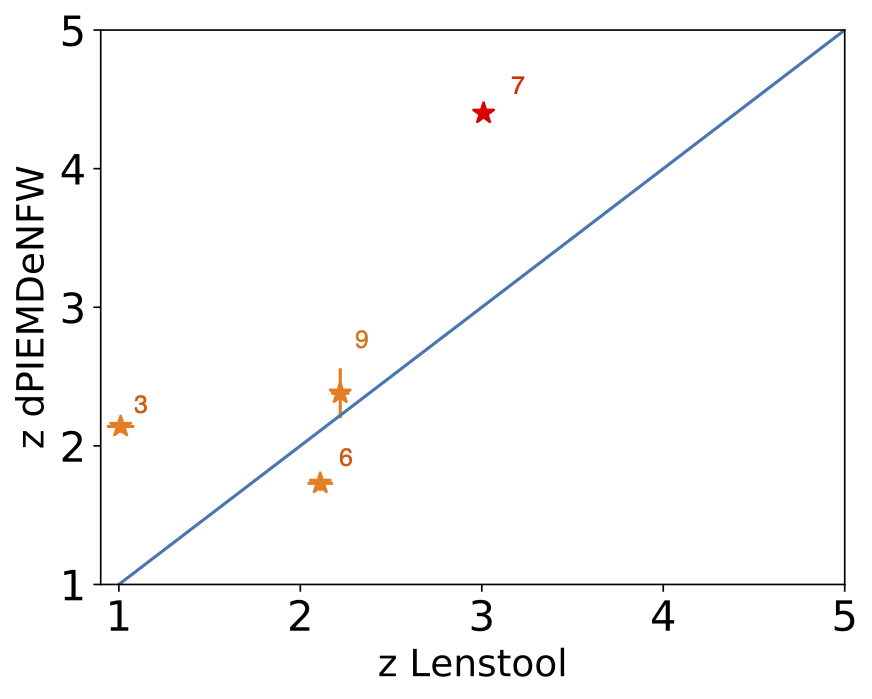}
     \includegraphics[width=0.32\linewidth]{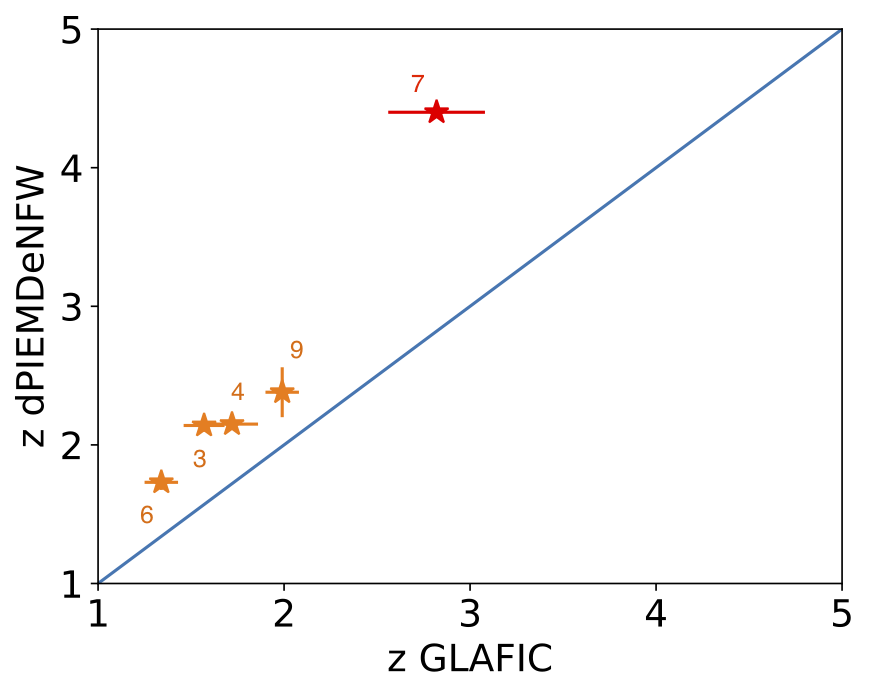}
     \includegraphics[width=0.32\linewidth]{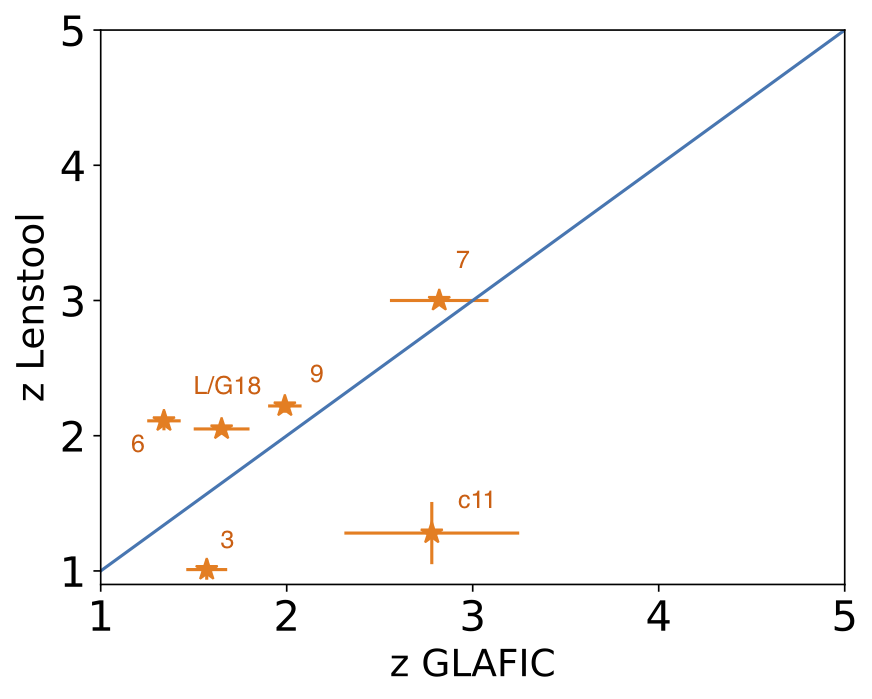}   
    \caption{Comparison of the redshift estimates between the different models presented in this work for the multiple images without a spectroscopic redshift confirmation. A line with a slope of unity is plotted in blue to guide the eye. System 7, which is considered with a fixed redshift of $zphot=4.4$ in the LTM and dPIEeNFW models is plotted in red. Note that the statistical uncertainties (shown as the $1\sigma$ uncertainties from the MCMC sampling) do not encompass the differences between models, that are more representative of the underlying systematic uncertainties.}
    \label{zcomp}
\end{figure*}

Finally, we compare in Figure \ref{zcomp} the model-predicted redshifts among the four models presented in this work, which are also reported in Tables \ref{tab:MI} and \ref{tab:zcomp}. We find that the LTM and dPIEeNFW models are in good agreement with each other, however they predict systematically higher redshifts than the Lenstool and GLAFIC models. The redshifts obtained with the Lenstool pipeline are slightly overestimated with respect to those derived with GLAFIC but remain in fairly good agreement. Part of the reason for the systematic underestimate of the Lenstool and GLAFIC models compared to the LTM and dPIEeNFW models is that the highest-redshift multiple image system, system 7, was fixed to its dropout redshift value for the LTM and dPIEeNFW models while in the Lenstool and GLAFIC models it was not, resulting in systematically lower redshifts. It should also be noted that the statistical uncertainties do not encompass the differences between models, that are more representative of the underlying uncertainties.

Note that all the SL models presented in this work have been built with only one multiply imaged system that has a spectroscopic redshift confirmation (with its two emission knots used as constraints in the modeling). It will be interesting to revise the differences between the models when more secure redshifts are measured. 

\section{The lensed star-forming galaxy at z$\sim3.6$}\label{sec:SYS1}

The lensed star-forming galaxy at z=3.6314 (corresponding to our systems 1 and 2) was studied by \citet{Dessauges2017}. They used the Lenstool pipeline to constrain the mass distribution of RXC0032 lensing model based on the z=3.6314 system (considering the two emission knots as in the other models presented in this work), as well as two other, triply imaged systems which were spectroscopically confirmed (although the redshift was not published). They derived the magnification for the lensed images of systems 1 and 2, as shown in Table \ref{magcomp}, which are necessary to study the intrinsic properties of the source.

Here, we compare in Table \ref{magcomp} the magnification estimated for each image of the system derived by the LTM, dPIEeNFW, Lenstool, GLAFIC algorithms and the model presented in \citet{Dessauges2017}. The magnification values between the strong lens models differ significantly (the dispersion between models being larger than the statistical uncertainties), which makes a reliable estimation of the intrinsic properties of the background source challenging. It will be interesting, when more spectroscopic data becomes available, to analyse if the lens models converge toward similar magnification values.

\begin{deluxetable*}{c c c c c c c}
\tablecaption{\label{tab:sys1}
    Magnification estimates for the lensed star-forming galaxy at z $\sim 3.6$}
\tablehead{Pipeline & 1.1 $\&$ 2.1 & 1.2 $\&$ 2.2 &1.3 $\&$ 2.3 &1.4 $\&$ 2.4 &1.5 $\&$ 2.5 & 1.6 $\&$ 2.6}
\startdata
LTM  & 9.5 $\pm$ 0.4 & 3.4 $\pm$ 0.4 & 26.5 $\pm$ 2.0 & 15.5 $\pm$ 1.5 & 5.0 $\pm$ 0.2 & 0.6 $\pm$ 0.1\\ 
dPIEeNFW  & 9.7 $\pm$ 0.7 & 9.4 $\pm$ 0.7 & \emph{unconstrained} & 11.5 $\pm$ 1.0 & 7.2 $\pm$ 0.5 & 4.4 $\pm$ 0.2\\ 
Lenstool & 17.2$\pm$1.3 & 9.0$\pm$0.9 & 46.6 $\pm$ 4.5& 35.0 $\pm$3& 9.7 $\pm$0.7& 3.8$\pm$0.4 \\ 
GLAFIC & 7.3 $\pm$ 1.1 &  3.4 $\pm$ 0.5 & 18.4 $\pm$ 3.5 & 10.8 $\pm$ 2.2 & 4.1 $\pm$ 0.6 & 1.4 $\pm$ 0.2\\ 
\citep{Dessauges2017}\tablenotemark{a} & 10.8 $\pm$ 4& 12.6 $\pm$ 3 & 24.7 $\pm$ 2 & 12.7 $\pm$ 3  & - & -\\ 
\hline
 Standart deviation\tablenotemark{b} & 1.6  & 4.0 & 53.0  & 10.2 &2.5  &  1.8\\ 
\enddata
\tablenotetext{a}{The strong lensing is obtained with the Lenstool pipeline using systems 1 and 2 as constraints as well as two other systems with spectroscopic redshifts measurements.}
\tablenotetext{b}{Standart deviation of the magnification values between the different models.}
\label{magcomp}
\end{deluxetable*}

\section{Summary}\label{sec:DIS}
The RELICS survey was designed to efficiently discover and characterize bright high-redshift galaxies that are magnified by massive cluster lenses, as well as to identify which galaxy clusters are the most efficient lenses for future follow-up campaigns \citep{Coe2019}. 
Based mostly on RELICS observations, we present here a full SL analysis of the merging galaxy cluster RXC J0032.1+1808. More recently, efforts have focused on SL systematic uncertainties arising from different modeling techniques \citep{Johnson2016, Meneghetti2017, Remolina2018}. 
In this work, we have adopted two different methodologies, the LTM technique and a fully parametric model, dPIEeNFW. We have also compared our results with the models obtained with the Lenstool and GLAFIC pipelines, that were independently constructed, so that our conclusions can be made more robust. As we show throughout, the results for most quantities of interest seem to agree fairly well between the different models. In that sense, differences between the LTM, dPIEeNFW, Lenstool and GLAFIC resulting models are then more representative of the true underlying uncertainties, than the magnitude of the statistical uncertainties from the respective minimization procedures.

The derived mass distribution and Einstein radius of RXC0032 reveals a very prominent lens, with a large effective Einstein radius of $\theta_E\sim40\arcsec$ at $z_s=2.0$, as supported by all models probed here. Since mergers enhance the lensing cross-section, merging clusters such as RXC0032 are of particular interest for the statistical study of the strongest gravitational lenses and are particularly useful when comparing Einstein radius distributions from observations to those from theoretical expectations \citep{Redlich2012, Redlich2014}.
While RELICS has only uncovered three high-redshift galaxy candidates in this field \citep{Salmon2020}, we find that RXC0032 is a promising lens to carry out wider and deeper imaging follow-up in order to expand the present coverage to all high-magnification regions.

All the lens models presented in this work and their corresponding deflection fields, magnification maps for different redshifts as well as a hundred random models to calculate errors, are made publicly available through the MAST archive\textsuperscript{\ref{mast}}. \\

This work is based on observations taken by the RELICS Treasury Program (GO-14096) with the NASA/ESA \textit{HST}. Program GO-14096 is supported by NASA through a grant from the Space Telescope Science Institute, which is operated by the Association of Universities for Research in Astronomy, Inc., under NASA contract NAS5-26555. 
This work was performed in part under the auspices of the U.S. Department of Energy by Lawrence Livermore National Laboratory under Contract DE-AC52-07NA27344.
This work was supported in part by World Premier International Research Center Initiative (WPI Initiative), MEXT, Japan, and JSPS KAKENHI Grant Number 
JP15H05892 and JP18K03693.
K.U. acknowledges support from the Ministry of Science and Technology of Taiwan under the grant MOST 106-2628-M-001-003-MY3. 
This work is partially supported by the Australian Research Council Centre of Excellence for All-Sky Astrophysics in 3 Dimensions (ASTRO-3D). 
S.T. acknowledges support from the ERC Consolidator Grant funding scheme (project ConTExt, grant No. 648179). The Cosmic Dawn Center is funded by the Danish National Research Foundation.

\facilities{HST, Spitzer, MAST}




\bibliographystyle{aasjournal}
\bibliography{ref}



\appendix

\section{List of multiple images and candidate identifications for RXC J0032.1+1808}

\begin{longrotatetable}
\begin{deluxetable*}{c c c c c c c c c c c}
\label{tab:MI}
\tablecaption{Multiple images and candidate identifications for RXC J0032.1+1808.}
\tablehead{Arc ID\tablenotemark{a} & R.A. & Decl& $\mathrm{z_{phot}}$ [$\mathrm{z_{min}}$-$\mathrm{z_{max}}$]\tablenotemark{b} & $\mathrm{z_{spec}}$ & $\mathrm{z_{model}^{LTM}}$[$68\%$ C.I.]\tablenotemark{c}& $\mathrm{z_{model}^{dPIEeNFW}}$[$68\%$ C.I.]\tablenotemark{c}&LTM/ & Lenstool\tablenotemark{d} &GLAFIC\tablenotemark{d} &RMS\tablenotemark{e}\\  
&[deg]&[deg]&&&&&dPIEeNFW\tablenotemark{d}&&&[$\arcsec$]}
\startdata
1.1  & 8.031533& 18.114345  &0.24 [0.17-0.64] &3.6314\tablenotemark{f}&-& -& $\checkmark$&$\checkmark$&$\checkmark$& 1.32\\ 
1.2  & 8.032142 & 18.113763  & 0.37 [0.27-0.43]&"&- &-&$\checkmark$ &$\checkmark$&$\checkmark$&0.40\\  
1.3  & 8.032671 & 18.112831   & 0.31 [0.23-0.47]&"&- &-&$\checkmark$ &$\checkmark$&$\checkmark$&1.58\\   
1.4  & 8.034162 & 18.111284   & 3.99 [0.25-4.36]&"&- &-&$\checkmark$ &$\checkmark$&$\checkmark$&0.65\\  
1.5   & 8.040679& 18.106825    &0.44 [0.24-0.69] &"&- &-&$\checkmark$ &$\checkmark$&$\checkmark$&2.15\\  
1.6  & 8.031840 & 18.113526   & -&"&-&- &$\checkmark$ &&$\checkmark$&0.34\\  
\hline
2.1  & 8.031454 & 18.114526   &3.54 [3.42-3.65]  &3.6314\tablenotemark{f}&-& -&&&&1.36 \\  
2.2  & 8.032260 & 18.113637    & 4.01 [3.90- 4.11]&"&- &-& $\checkmark$&$\checkmark$&$\checkmark$&0.70\\  
2.3   & 8.032492 & 18.113161   & 3.61 [3.55-3.74]&"&-&- &$\checkmark$&$\checkmark$&$\checkmark$&1.15 \\  
2.4   & 8.034382 & 18.111196   & 3.61 [3.51-3.77]&"&-&- &$\checkmark$&$\checkmark$&$\checkmark$&0.44 \\  
2.5  & 8.040661 & 18.106926   & 0.30 [0.11-3.61]&"&-&- & $\checkmark$&$\checkmark$&$\checkmark$&2.06\\   
2.6   & 8.031893 & 18.113509   &- &"&-&- &$\checkmark$&&$\checkmark$&0.76\\ 
\hline
3.1 & 8.050767 & 18.131576   &1.05 [1.01-1.06] &-&2.10 [2.04-2.20]&  2.14 [2.14-2.23] &$\checkmark$&$\checkmark$&$\checkmark$& 0.49\\   
3.2 &  8.049925 & 18.131756   &1.05 [1.01-1.76] &-&&&$\checkmark$&$\checkmark$&$\checkmark$&2.50 \\ 
3.3 & 8.035982  & 18.134820   & 1.63 [1.00-1.79] &-&& &$\checkmark$&&$\checkmark$&0.83\\
\hline
4.1 & 8.051190 & 18.131482   & -&-& 2.23 [2.00-2.21] &  2.15 [2.08-2.16]&$\checkmark$&&$\checkmark$&1.04 \\   
4.2 &  8.049535 & 18.131823   &- &-&& &$\checkmark$&&$\checkmark$&1.86 \\ 
4.3 &  8.036149 & 18.134948  & 0.97 [0.70-1.96] &-&& &$\checkmark$&&$\checkmark$& 1.62 \\
\hline
c5.1 & 8.051015 &18.131285 &1.70 [1.58-1.78]&-&$\sim2.3$&$\sim2.0$&&&$\checkmark$&-\\ 
c5.2 & 8.049483 &18.131599 & 1.41 [1.39-1.81]&-&&&&&$\checkmark$&-\\
c5.3 & 8.035952 & 18.134634 &1.59 [1.13-1.92] &-&&&&&$\checkmark$&-\\   
\hline
6.11 & 8.042302 &  18.142446  & 1.71 [1.71-1.78]&-& 1.61 [1.56-1.70]& 1.73 [1.73-1.84]&$\checkmark$&$\checkmark$&$\checkmark$&0.24\\   
6.12 & 8.046450& 18.140492   &1.10 [1.02-1.14] &-&& &$\checkmark$&$\checkmark$&$\checkmark$& 0.62\\   
6.13 & 8.059917& 18.139378   &  1.11 [1.02-1.71] &-&& &$\checkmark$&$\checkmark$&$\checkmark$&1.65 \\   
\hline
6.21 & 8.042491& 18.142400 & 1.03 [0.93-1.09] &-&1.64 [1.53-1.60] &1.89 [1.78-1.92] &$\checkmark$ &&$\checkmark$&0.26\\   
6.22 & 8.046380 & 18.140578 & -&-&& &$\checkmark$&&$\checkmark$& 0.69\\   
6.23 &8.060023 & 18.139325 & - &-&& & $\checkmark$&&$\checkmark$&1.70 \\   
\hline
7.1 & 8.036775 & 18.128377   &4.41 [0.42-4.66] &-& 4.4\tablenotemark{g} & 4.4\tablenotemark{g}&$\checkmark$&$\checkmark$&$\checkmark$&1.25 \\   
7.2 & 8.043467 & 18.126995   &  0.30 [0.16-3.96]&-& &&$\checkmark$ &$\checkmark$&$\checkmark$&2.13\\ 
7.3 &  8.060019&  18.123190 &  &-& &&$\checkmark$&$\checkmark$&$\checkmark$& 3.15\\ 
\hline
c8.1 & 8.051712& 18.122845   & 2.82 [2.74-3.11]&-&$\sim2.7$ & $\sim2.7$&&&$\checkmark$& -\\   
c8.2 & 8.049750 & 18.123512   & 0.05 [0.02-3.07]&-& &&&&$\checkmark$& -\\
\hline
9.1 & 8.047754&18.114628   &3.25 [3.00-3.53] &-& 2.83 [2.84-3.06] & 2.38 [2.03-2.39] &$\checkmark$&$\checkmark$&$\checkmark$& 3.70\\   
9.2 & 8.053467& 18.117267   & -&-& &&$\checkmark$&$\checkmark$&$\checkmark$& 2.13\\ 
9.3 &8.032281  & 18.122876  & 2.14 [1.85-2.28] &-& &&$\checkmark$&$\checkmark$&$\checkmark$& 2.39\\ 
\hline
c10.1 & 8.053825 & 18.117562   &  2.32 [2.17-2.53]&-& $\sim2.7$ & $\sim2.2$ &&&& -\\   
c10.2 & 8.047263 & 18.114624   & -&-&& &&&& -\\  
c10.3 &  8.032309&  18.123215  & -&-&& & &&&-\\
\hline
c11.1 & 8.049167 & 18.153004   &2.76 [0.07-3.08] &-& $\sim3$ & $\sim3$ &&$\checkmark$&$\checkmark$& -\\   
c11.2  & 8.047482 & 18.152860 & 2.71 [0.36-3.10]  &-& &&&$\checkmark$&$\checkmark$& -\\ 
\hline
c12.1 & 8.049383 &  18.152779  & -&-&$\sim3$ & $\sim3$&&&& -\\   
c12.2 & 8.048008 & 18.152721  &  1.36 [0.59-2.66] &-&& &&&& -\\ 
\hline
c13.1 & 8.048654 &  18.152527  & -&-&$\sim3$ & $\sim3$&&&& -\\   
c13.2 & 8.047547 &  18.152442 &1.54 [0.16-2.98] &-&& &&&& -\\ 
\hline\hline
L14.1 & 8.046523 &  18.155275  & 0.55 [0.46-4.54]&-& -& -&&$\checkmark$&& -\\   
L14.2 & 8.051645 &  18.155275 & 0.92 [0.75-1.42]&-&-&- &&$\checkmark$&& -\\ 
\hline
L15.1 & 8.043844 &  18.148375  & 3.11 [0.26-3.32]&-&- &- &&$\checkmark$&& -\\   
L15.2 & 8.061134 &  18.144614 & 3.19[1.87-3.27]&-&-&- &&$\checkmark$&& -\\ 
\hline
L16.1 & 8.054572 &  18.140649  & -&-&- &- &&$\checkmark$&& -\\   
L16.2 & 8.051372 &  18.140131 & -&-&-& -&&$\checkmark$&& -\\ 
\hline
L17.1 & 8.040382 & 18.131975 & 4.23 [0.34-4.49]&-&- &-&&$\checkmark$&& -\\
L17.2 & 8.042291 & 18.131664 & 0.77 [0.19-0.90]&-& -&-&&$\checkmark$&& -\\
\hline
L/G18.1 & 8.039637 & 18.131971   &2.64 [2.35-2.92] &-& -&-&&$\checkmark$&$\checkmark$& -\\
L/G18.2 & 8.043413 & 18.131387   & 2.27 [2.22-2.55]&-&- &-&&$\checkmark$&$\checkmark$& -\\
L18.3 & 8.059395 & 18.127467   & 2.31 [2.04-2.61]&-&- &-&&$\checkmark$&& -\\
\hline
L19.1 & 8.051801 & 18.126669   &  3.94 [3.46-4.30]&-& -&-&&$\checkmark$&& -\\
L19.2 & 8.052461 & 18.12658   & 4.37 [4.24-4.48] &-& -&-&&$\checkmark$&& -\\
L19.3 & 8.028309 & 18.131135   &3.87 [3.06-4.50] &-& -&-&&$\checkmark$&& -\\
\hline
L20.1 & 8.050102 & 18.123303   & 2.43 [0.15-2.90]&-& -&-&&$\checkmark$&& -\\
L20.2 & 8.051282 & 18.122912   & 0.08 [0.05-3.08]&-&- &-&&$\checkmark$&& -\\
\hline
G21.1 & 8.048514 & 18.107541   &  1.59 [0.45-2.57]&-& -&-&&&$\checkmark$& -\\
G21.2 & 8.047208 & 18.107382   & 1.58 [0.46-2.55]&-& -&-&&&$\checkmark$& -\\
\hline
L22.1 & 8.035258 & 18.106743   &- &-&- &-&&$\checkmark$&& -\\
L22.2 & 8.033758 & 18.107896   &- &-&- &-&&$\checkmark$&& -\\
\hline
\enddata
\tablenotetext{a}{Identification of all multiple images uncovered. Those identified with a "c" refer to less reliable therefore candidate identifications that were not included in our SL models. Systems identified with an "L" or "G" refer to those independently identified by the Lenstool and GLAFIC modeling teams, respectively.}
\tablenotetext{b}{Photometric redshift with upper and lower limits, based on the BPZ estimates from the RELICS catalog with the $95\%$ confidence range; a "-" sign indicates an image for which its $\mathrm{z_{phot}}$ could not be measured due to light contamination or poor signal-to-noise ratio.}
\tablenotetext{c}{Redshift prediction based on our LTM and analytical dPIEeNFW best-fit models, respectively.}
\tablenotetext{d}{The symbol $\checkmark$ indicates whether a multiple image is used in the SL modeling by the LTM/dPIEeNFW, Lenstool and GLAFIC pipelines.}
\tablenotetext{e}{RMS between the observed and model-predicted multiple images from our LTM best-fit model.}
\tablenotetext{f}{Measured by \citet{Dessauges2017}, considered as a fixed redshift in our SL models.}
\tablenotetext{g}{We fix the redshift of system 7 which is a 'dropout' object.}
\end{deluxetable*}
\end{longrotatetable}

\startlongtable
\begin{deluxetable*}{c c c c c}
\tablecaption{\label{tab:zcomp}
    Redshift estimates for the multiple images in the RXC J0032.1+1808 cluster field from the Lenstool and GLAFIC models.}
\tablehead{Arc ID\tablenotemark{a} & $\mathrm{z_{model}^{Lenstool}}$[$68\%$ C.I.]\tablenotemark{b} & $\mathrm{z_{model}^{GLAFIC}}$[$68\%$ C.I.]\tablenotemark{b} &$RMS_{Lenstool}$\tablenotemark{c}&$RMS_{GLAFIC}$\tablenotemark{c}\\  
&&&[$\arcsec$]&[$\arcsec$]}
\startdata
1.1  &3.6314\tablenotemark{d}&3.6314\tablenotemark{d}&0.55& 0.57\\ 
1.2  &"&"&0.32&0.05\\  
1.3  &"&"&0.41&0.33\\   
1.4  &"&"&0.49&0.54\\  
1.5  &"&"&0.75&0.96\\  
1.6  &"&"&-&0.12\\  
\hline
2.1  &3.6314\tablenotemark{d}&3.6314\tablenotemark{d}&0.56&0.47 \\  
2.2  &"&"&0.38&0.03\\  
2.3  &"&"&0.05&0.43\\  
2.4  &"&"&0.58&0.52\\  
2.5  &"&"&0.62&1.07\\   
2.6  &"&"&-&0.19\\ 
\hline
3.1 &1.01 [1.00-1.15]&1.57 [1.47-1.69]&0.11&0.28\\   
3.2 &"&"&0.17&1.14\\ 
3.3 &-&"&-&0.30\\
\hline
4.1 &-& 1.72 [1.57-1.85]&-&0.38 \\   
4.2 &-&"&-& 0.36\\ 
4.3 &-&-&&- \\
\hline
c5.1&-&1.68 [1.55-1.88]&-&0.33\\ 
c5.2 &-&"&-&0.31\\
c5.3 &-&-&-&-\\   
\hline
6.11 &2.11 [2.05-2.19]&-&0.60&-\\   
6.12 &"&-&0.94&- \\   
6.13 &"&-&0.09& -\\   
\hline
6.21 &-&1.34 [1.29-1.47]&-&0.58\\   
6.22 &-&"&-&0.38\\   
6.23 &-&"&-& 0.19\\   
\hline
7.1 &3.00 [3.00-3.01] &2.82 [2.54-3.07]&0.90&0.91\\   
7.2 &"&"&0.31&0.26\\ 
7.3 &"&"&0.93&0.86\\ 
\hline
c8.1 &-&2.83 [2.63-3.12]&0.50&0.15\\   
c8.2 &-&"&0.40&0.15\\
\hline
9.1 &2.22 [2.16-2.26]&1.99  [1.91-2.09]&0.21&0.50\\   
9.2 &"&"&0.60&0.58\\ 
9.3 &"&"&0.80&0.13\\ 
\hline
c11.1 &1.28 [1.22-1.68]&2.78 [2.28-3.22]&0.22&0.13\\   
c11.2&-&"&0.25&0.04\\ 
\hline
L14.1 & 1.95 [1.74-3.52] &-&0.31&-\\   
L14.2 &"&-&0.08&-\\
\hline
L15.1  &3.01 [3.01-3.05] &-&0.53&-\\
L15.2  &"&-&1.22&-\\
\hline
L16.1  &2.43 [2.24-2.97] &-&0.62&-\\   
L16.2  &"&-&0.73&-\\ 
\hline
L17.1  &2.54 [2.40-2.66] &-&0.54&-\\
L17.2  &"&-&0.47&-\\
\hline
L/G18.1 &2.05 [2.00-2.08] &1.65 [1.51-1.81]&0.60&0.37\\
L/G18.2 &"&"&0.98&0.30\\
L18.3 &"&-&0.63&-\\
\hline
L19.1  & 5.00 [4.97-5.00] &-&0.64&-\\
L19.2  &"&-&0.48&-\\
L19.3  &"&-&1.19&-\\
\hline
L20.1 &2.70 [2.58-2.85] &-&0.48&-\\
L20.2  &"&-&0.47&-\\
\hline
G21.1 &-& 1.74 [1.58-1.97]&-&0.03\\
G21.2 &-&"&-&0.02\\
\hline
L22.1  &1.43 [1.33-1.55]&-&0.08&-\\
L22.2  &"&-&0.06&-\\
\hline
\enddata
\tablenotetext{a}{Identification of the multiple images following Figure \ref{cc} and Table \ref{tab:MI}.}
\tablenotetext{b}{Redshift prediction based on Lenstool and GLAFIC best-fit models with the $1\sigma$ statistical uncertainty, respectively.}
\tablenotetext{c}{Individual RMS between the observed and model-predicted multiple images from the Lenstool and GLAFIC models.}
\tablenotetext{d}{Considered as a fixed redshift in the Lenstool and GLAFIC models.}
\end{deluxetable*}

\section{Reproduction of multiple images and candidate identifications for RXC J0032.1+1808}

\begin{figure*}
    \centering
    \includegraphics[width=\linewidth]{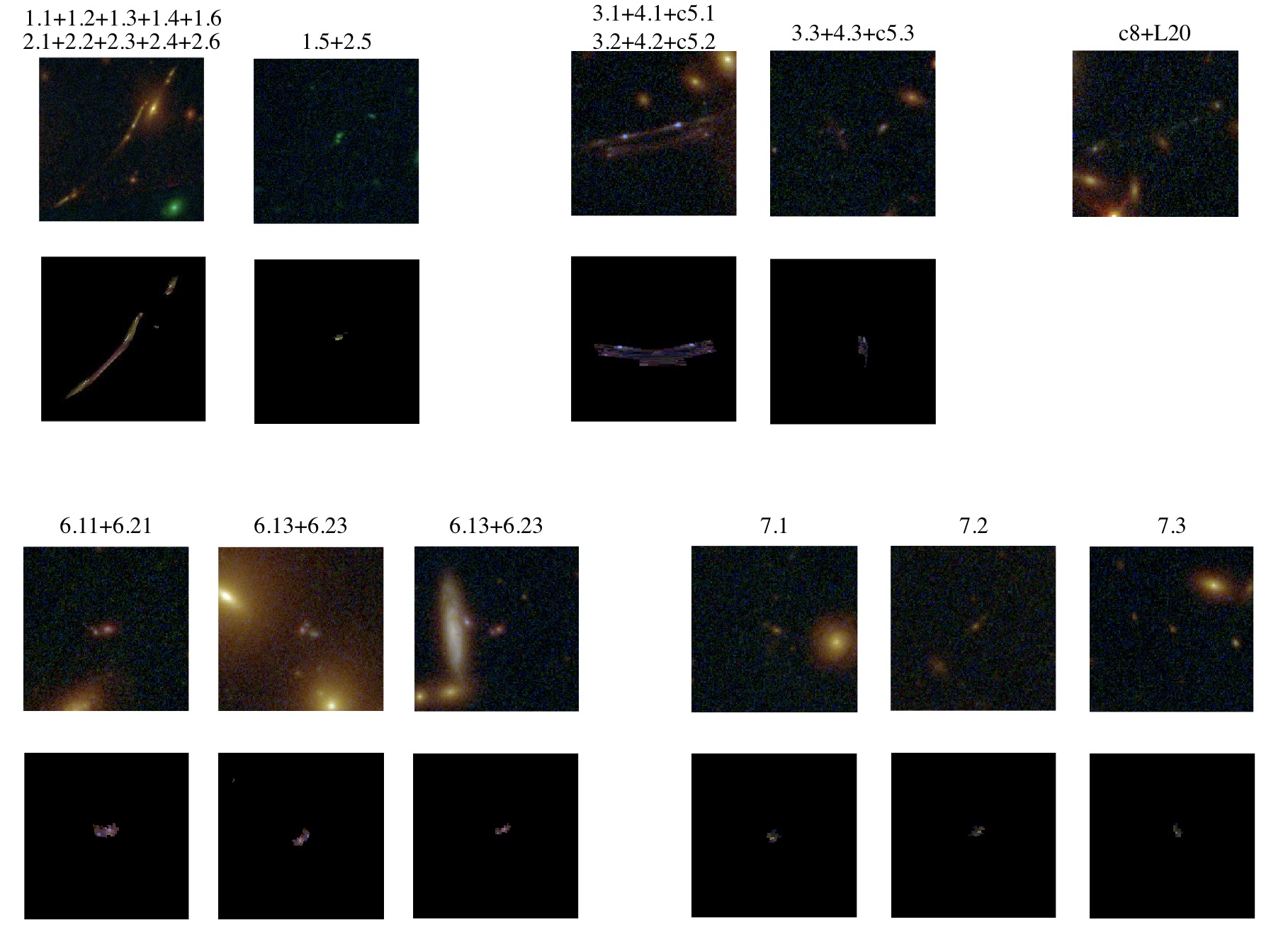} \hfill\\
    \includegraphics[width=\linewidth]{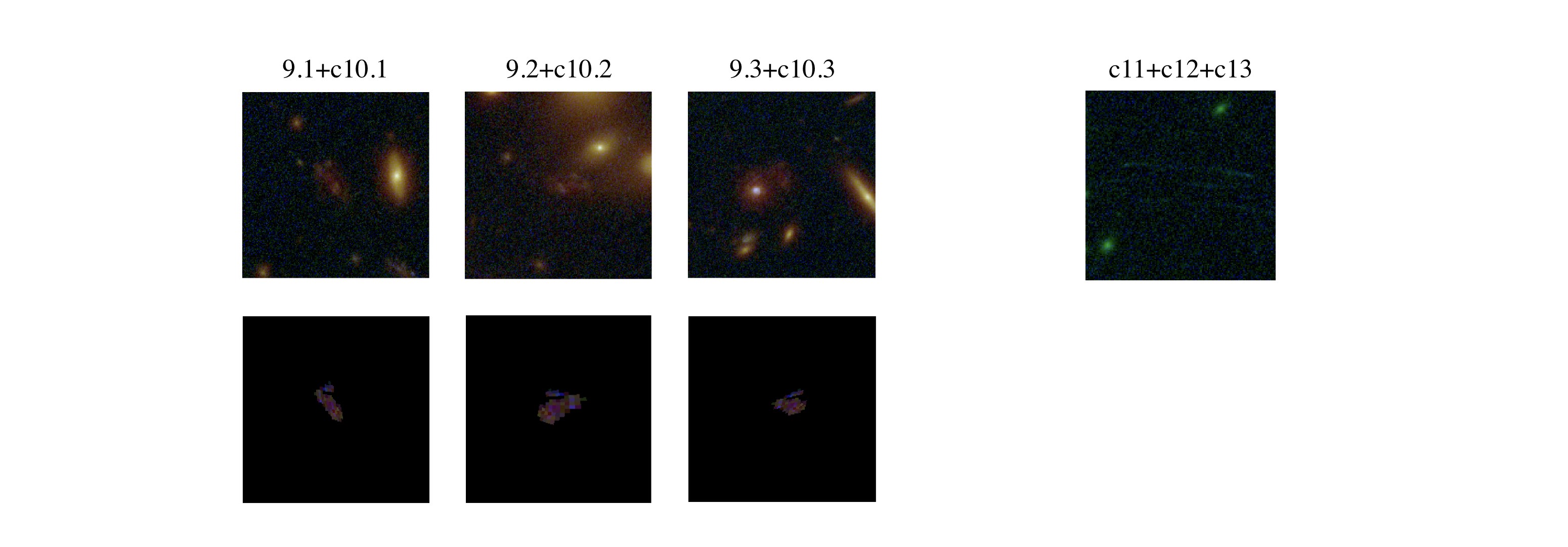}
    \caption{Reproduction of the multiple images used in the minimisation process with our best-fit LTM model. For each image, we de-lens the first image of the system to the source plane and back to the image plane to compare to the other images of that system, finding that both the orientation and internal details of the model-predicted images (bottom rows) are similar to those of the observed images (upper rows). The reconstructions have been manually centered on the observed positions and thus do not portray the modeled positions. }
    \label{st1}
\end{figure*} 

\begin{figure*}
    \centering
    \includegraphics[width=\linewidth]{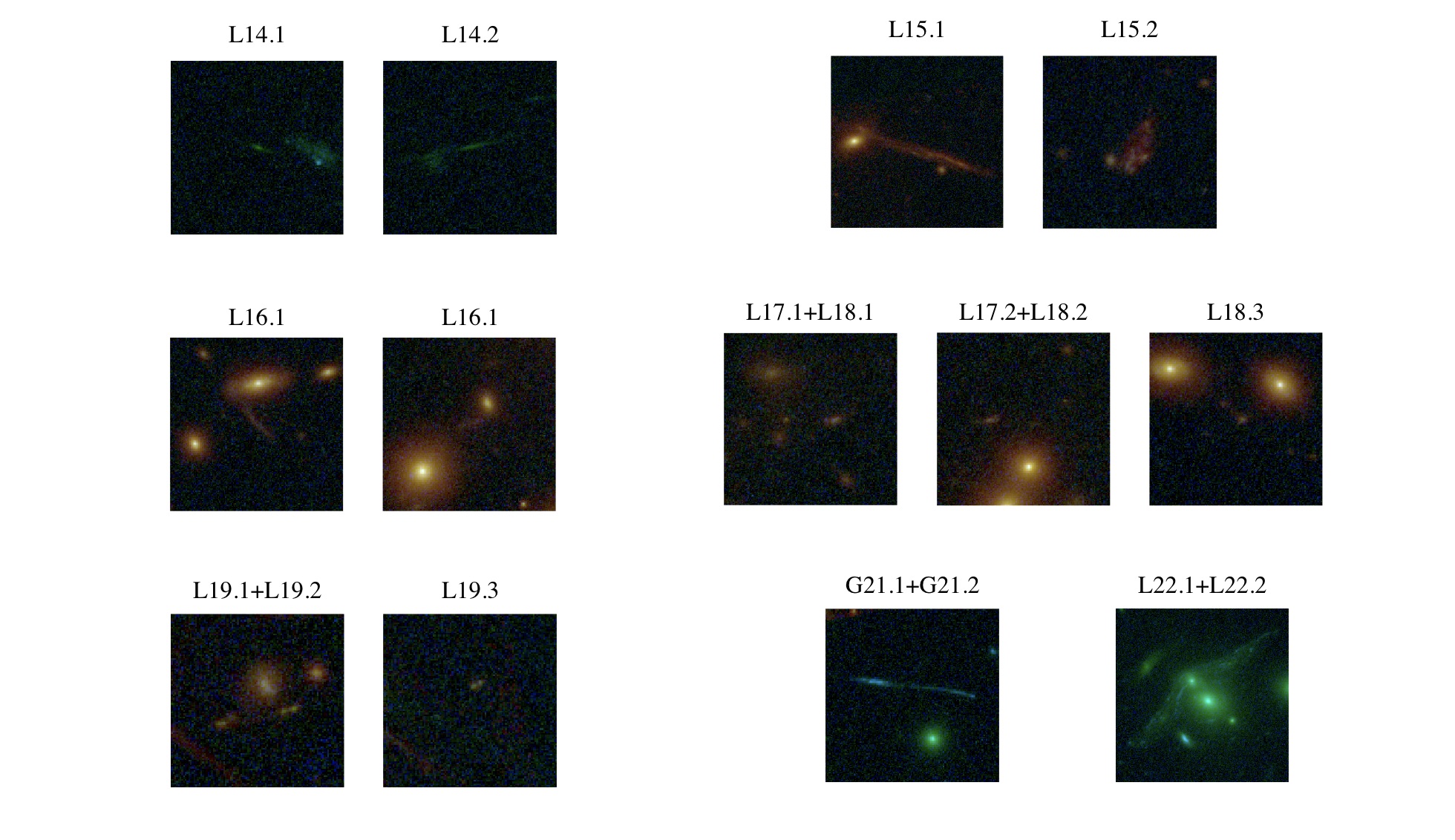}
    \caption{HST cut-outs of the multiple images to illustrate the systems independently identified by the Lenstool or GLAFIC teams.}
    \label{st2}
\end{figure*} 

\end{document}